\documentclass[twocolumn]{aastex631}

\received{\today}

\submitjournal{ApJ}

\shortauthors{Costantin et al.}

\begin{document}

\title{Expectations of the size evolution of massive galaxies at $3 \leq z \leq 6$ from the TNG50 simulation: \\
the CEERS/JWST view}

\suppressAffiliations

\correspondingauthor{Luca Costantin}
\email{lcostantin@cab.inta-csic.es}


\collaboration{20}{}
\author[0000-0001-6820-0015]{Luca Costantin}
\affiliation{Centro de Astrobiolog\'{\i}a (CAB), CSIC-INTA, Ctra. de Ajalvir km 4, Torrej\'on de Ardoz, E-28850, Madrid, Spain}

\author[0000-0003-4528-5639]{Pablo G.~P\'erez-Gonz\'alez}
\affiliation{Centro de Astrobiolog\'{\i}a (CAB), CSIC-INTA, Ctra. de Ajalvir km 4, Torrej\'on de Ardoz, E-28850, Madrid, Spain}

\author[0000-0003-2338-5567]{Jes\'us Vega-Ferrero}
\affiliation{Instituto de Astrof\'isica de Canarias, 38200, La Laguna, Tenerife, Spain}

\author[0000-0002-1416-8483]{Marc Huertas-Company}
\affiliation{Instituto de Astrof\'isica de Canarias, 38200, La Laguna, Tenerife, Spain}
\affiliation{Departamento de Astrof\'isica, Universidad de La Laguna, 38205, La Laguna, Tenerife, Spain}
\affiliation{LERMA, Observatoire de Paris, CNRS, PSL, Universit\'e de Paris, France}

\author[0000-0003-0492-4924]{Laura Bisigello}
\affiliation{Dipartimento di Fisica e Astronomia "G.Galilei", Universit\'a di Padova, Via Marzolo 8, I-35131 Padova, Italy}
\affiliation{INAF--Osservatorio Astronomico di Padova, Vicolo dell'Osservatorio 5, I-35122, Padova, Italy}

\author[0000-0002-2861-9812]{Fernando Buitrago}
\affiliation{Departamento de F\'{i}sica Te\'{o}rica, At\'{o}mica y \'{O}ptica, Universidad de Valladolid, 47011 Valladolid, Spain}
\affiliation{Instituto de Astrof\'{i}sica e Ci\^{e}ncias do Espa\c{c}o, Universidade de Lisboa, OAL, Tapada da Ajuda, PT1349-018 Lisbon, Portugal}

\author[0000-0002-9921-9218]{Micaela B. Bagley}
\affiliation{Department of Astronomy, The University of Texas at Austin, Austin, TX, USA}

\author[0000-0001-7151-009X]{Nikko J. Cleri}
\affiliation{Department of Physics and Astronomy, Texas A\&M University, College Station, TX, 77843-4242 USA}
\affiliation{George P.\ and Cynthia Woods Mitchell Institute for Fundamental Physics and Astronomy, Texas A\&M University, College Station, TX, 77843-4242 USA}

\author[0000-0003-1371-6019]{Michael C. Cooper}
\affiliation{Department of Physics \& Astronomy, University of California, Irvine, 4129 Reines Hall, Irvine, CA 92697, USA}

\author[0000-0001-8519-1130]{Steven L. Finkelstein}
\affiliation{Department of Astronomy, The University of Texas at Austin, Austin, TX, USA}

\author[0000-0002-4884-6756]{Benne W. Holwerda}
\affil{Physics \& Astronomy Department, University of Louisville, 40292 KY, Louisville, USA}

\author[0000-0001-9187-3605]{Jeyhan S. Kartaltepe}
\affiliation{Laboratory for Multiwavelength Astrophysics, School of Physics and Astronomy, Rochester Institute of Technology, 84 Lomb Memorial Drive, Rochester, NY 14623, USA}

\author[0000-0002-6610-2048]{Anton M. Koekemoer}
\affiliation{Space Telescope Science Institute, 3700 San Martin Dr., Baltimore, MD 21218, USA}

\author[0000-0001-8421-5890]{Dylan Nelson}
\affiliation{Zentrum f\"ur Astronomie der Universit\"at Heidelberg, ITA, Albert-Ueberle-Str. 2, D-69120 Heidelberg, Germany}

\author[0000-0001-7503-8482]{Casey Papovich }
\affiliation{Department of Physics and Astronomy, Texas A\&M University, College Station, TX, 77843-4242 USA}
\affiliation{George P.\ and Cynthia Woods Mitchell Institute for Fundamental Physics and Astronomy, Texas A\&M University, College Station, TX, 77843-4242 USA}

\author[0000-0003-1065-9274]{Annalisa Pillepich}
\affiliation{Max-Planck-Institut fur Astronomie, K\"onigstuhl 17, 69117 Heidelberg, Germany}

\author[0000-0003-3382-5941]{Nor Pirzkal}
\affiliation{ESA/AURA Space Telescope Science Institute}

\author[0000-0002-8224-4505]{Sandro Tacchella}
\affiliation{Kavli Institute for Cosmology, University of Cambridge, Madingley Road, Cambridge, CB3 0HA, UK}
\affiliation{Cavendish Laboratory, University of Cambridge, 19 JJ Thomson Avenue, Cambridge, CB3 0HE, UK}

\author[0000-0003-3466-035X]{L. Y. Aaron\ Yung}
\affiliation{Astrophysics Science Division, NASA Goddard Space Flight Center, 8800 Greenbelt Rd, Greenbelt, MD 20771, USA}


\begin{abstract}

We present a catalog of about 25,000 images of massive ($M_{\star} \ge 10^9 M_{\odot}$) galaxies at redshift $3 \leq z \leq 6$ from the TNG50 
cosmological simulation, tailored for observations at multiple wavelengths carried out with JWST. 
The synthetic images\footnote[1]{Data publicly released at \url{https://www.tng-project.org/costantin22}.} 
were created with the SKIRT radiative transfer code, including the effects of dust attenuation and scattering. 
The noiseless images were processed with the \texttt{mirage} simulator to mimic the Near Infrared Camera (NIRCam) observational strategy 
(e.g., noise, dithering pattern, etc.) of the Cosmic Evolution Early Release Science (CEERS) survey.
In this paper, we analyze the predictions of the TNG50 simulation for the size evolution of galaxies at $3 \leq z \leq 6$ and the expectations for CEERS to probe that evolution.
In particular, we investigate how sizes depend on wavelength, redshift, mass, and angular resolution of the images.
We find that the effective radius accurately describes the three-dimensional half-mass radius of TNG50 galaxies.
Sizes observed at 2~$\mu$m are consistent with those measured at 3.56~$\mu$m at all redshifts and masses.
At all masses, the population of higher-$z$ galaxies is more compact than their lower-$z$ counterparts.
However, the intrinsic sizes are smaller than the mock observed sizes for the most massive galaxies, especially at $z \lesssim 4$.
This discrepancy between the mass and light distribution may point to a transition in the galaxy morphology at $z$=4-5, 
where massive compact systems start to develop more extended stellar structures.

\end{abstract}




\section{Introduction \label{sec:section_1}}

The emergence of the Hubble sequence is one of the fundamental challenges 
of the hierarchical picture of galaxy assembly. 
In the local Universe, galaxies of late-type morphologies are star-forming systems dominated by the dynamical properties
of their stellar disk, while early-type galaxies are typically quiescent and exhibit spheroidal-dominated morphologies \citep{Lintott2011,Cappellari2016}.
On the contrary, the optical restframe morphology of more distant galaxies is increasingly irregular \citep{Driver1995, Papovich2005, Mortlock2013}
and the bulge and disk structures almost disappear at early cosmic time \citep{Ravindranath2006, HuertasCompany2016}.
Thus, why are galaxies different at different epochs? Do galaxies change their structure across time?
In the end, which are the morphological properties of the first galaxies?

In terms of their structure, there is a wide consensus, both from space- and ground-based observations, 
that galaxies experience a strong size evolution since $z=3$,
with galaxies at $z\sim2$ a factor 2-7 smaller in size than their local counterparts of similar masses
\citep{Daddi2005, Trujillo2007, Buitrago2008, Grogin2011, Koekemoer2011, vanderWel2014, Mowla2019, Suess2019, Mosleh2020}.
At higher redshift, a less steep evolution in size is found \citep{Oesch2010, CurtisLake2016},
but a fair census of high-$z$ galaxies, and the characterization
of their morphological structure, still has to come \citep[see e.g.,][]{Holwerda2020}.
While imaging carried out with the Hubble Space Telescope (HST)
has enabled the study of the optical rest-frame morphologies of galaxies up to $z\sim3$, 
the structural properties of higher-$z$ galaxies remain elusive.
Indeed, this task represents one of the main challenges that extragalactic observations carried out with
JWST are going to address, providing for the first time
the restframe optical and near-infrared morphology of galaxies at $z>3$
with unprecedented spatial resolution \citep[e.g., ][]{Kartaltepe2022, Kocevski2022, PerezGonzalez2022}.

In this context, the goal of this work is twofold: (1) calibrate how accurately morphology can be measured at redshift $3 \leq z \leq 6$;
(2) provide detailed predictions of state-of-the-art cosmological simulations to be compared with the first JWST observations.
To fulfill these goals, a powerful tool to investigate the synergy between observations and theory is 
the so-called \emph{forward modeling} of data, generating and analyzing
synthetic images from hydrodynamic simulations \citep[e.g., ][]{Snyder2015, Trayford2017, Wu2020, Popping2022}.
These noiseless images can be used to create mock observations
than mimic the observational strategy of any available facility,
providing crucial constraints on the expected performance of a given instrument.
Furthermore, such observations link the observed morphology of galaxies 
with underlying physical processes in powerful ways \citep{RodriguezGomez2019, Park2022}.

In this work, we describe the production of a mock image catalog
of high-$z$ galaxies ($3 \leq z \leq 6$) from the IllustrisTNG project 
\citep{Marinacci2018, Naiman2018, Nelson2018, Pillepich2018, Springel2018},
applying radiative transfer calculations to quantify the spatially resolved 
effects of dust on galaxy structures.
These synthetic images are created to mimic
noiseless observations both for NIRCam and MIRI instruments mounted on the JWST.  
Then, taking advantage of this dataset, we create and analyze mock observations for NIRCam,
the primary imager mounted on the JWST, which will
cover the infrared wavelength range 0.6 to 5 $\mu$m.
In particular, we aimed at testing the strategy and expected performances
of the Cosmic Evolution Early Release Science survey 
\citep[CEERS;][]{Finkelstein2017} in retrieving the morphology of high-$z$ galaxies
with observations at 2 and 3.56~$\mu$m.
Indeed, this catalog allows overcoming the limitations of mock observations
based on semi-analytic models currently available, 
providing a realistic and detailed description of the complex morphology of galaxies.
Furthermore, in a companion paper (Vega-Ferrero et al., \emph{in prep.}),
we use the same dataset to optimize a neural network model 
(based on the contrastive learning framework) to estimate galaxy's morphologies 
in an unsupervised and automated way.
 
The paper is organized as follows. In Sect.~\ref{sec:section2} we describe CEERS main goals
and the sample of galaxies selected from the TNG50 simulation. In Sect.~\ref{sec:section3}
we explain how synthetic noiseless images, raw data for NIRCam observations, and
fully calibrated images are created. In Sect.~\ref{sec:section4} we present and discuss the measured morphologies
at different wavelengths, redshifts, and masses.
Finally, we provide our conclusions in Sect.~\ref{sec:section5}.


\section{Sample and Data \label{sec:section2}} 

In this work, we model gas and star particles from the IllustrisTNG suite of simulations
to mimic NIRCam observations of high-$z$ galaxies following the observational
strategy of CEERS \citep[Finkelstein et al. (\emph{in prep.}), ][]{Bagley2022}.

\subsection{The CEERS Survey}

CEERS is an approved early-release science program of the JWST mission,
covering $\sim100$~arcmin$^2$ in the Extended Groth Strip field (EGS). 
CEERS is one of the first public JWST surveys with data publicly available.
The survey consists of 63 hours of NIRCam ($1-5~\mu$m) and MIRI ($5-21~\mu$m) imaging, 
NIRSpec $R\sim100$ and $R\sim1000$ spectroscopy, and NIRCam/grism $R\sim1500$ spectroscopy. 

In terms of the scientific deliverables, CEERS enables morphological studies with unprecedented
spatial resolution over a large wavelength range using NIRCam, well beyond the volume accessible to HST. 
In particular, it promises to unveil the details of early galaxy structures, 
including the formation of the first disks, the appearance of the first bulges, and the physical mechanisms
responsible for fueling and quenching star formation and active galactic nuclei.

In this work, we focus on the CEERS10 pointing strategy,
one of the ten CEERS pointing combining NIRCam and MIRI 
parallel observations, with MIRI as a primary instrument and NIRCam as a secondary one.
In particular, we replicate the observing mode of NIRCam observations,
which accounts for 9 groups per integration and a MEDIUM8 readout pattern.
We focus on the F200W and F356W bands, the most sensitive in the short and long wavelength regimes.
The planned exposure time is 2834 seconds, considering a three-point dithering pattern
optimized for parallel observations with MIRI F770W band. 
The designed $5\sigma$ limiting magnitudes for point-like sources are 28.97 mag and 28.95 mag
for the F200W and F356W filters, respectively.

\begin{deluxetable}{cccc}
\tablecaption{Properties of the sample galaxies from the TNG50 simulation.
\label{tab:table1}}
\tablehead{
\colhead{$z$} & \colhead{Number of galaxies} & \colhead{log($M_{\star}$)} & \colhead{log($M_{\star, \rm max}$)} \\
\colhead{} & \colhead{} & \colhead{$M_{\odot}$} &  \colhead{$M_{\odot}$} 
}
\decimalcolnumbers
\startdata
3--6 	&	1238	&	$9.39_{-0.30}^{+0.61}$  	&	11.48	\\
3 	& 	760	&	$9.43_{-0.33}^{+0.61}$  	&	11.48	\\
4 	& 	326	&	$9.33_{-0.25}^{+0.57}$  	&	11.24	\\
5	& 	113	&	$9.31_{-0.24}^{+0.56}$  	&	10.96	\\
6 	& 	39	&	$9.26_{-0.22}^{+0.59}$  	&	10.61	\\
\enddata
\tablecomments{
(1) Redshift of the simulation snapshot. (2) Number of galaxies. (3) Median stellar mass and 16th-84th percentile range.
(4) Maximum stellar mass.}
\end{deluxetable}

\subsection{The TNG50 cosmological simulation}

The IllustrisTNG Project \citep{Marinacci2018, Naiman2018, Nelson2018, Pillepich2018, Springel2018} 
is the follow-up of the original Illustris simulation \citep{Vogelsberger2014, Genel2014, Sijacki2015} and 
comprises three suites of magneto-hydrodynamic cosmological simulations (TNG300, TNG100, and TNG50)
evolved with the moving-mesh code AREPO \citep{Springel2010}.

Our analysis is based on TNG50-1 \citep[hereafter TNG50;][]{Pillepich2019, Nelson2019}, 
the latest and highest resolution version of IllustrisTNG simulation suite. 
TNG50 follows the evolution of $2 \times 2160^3$ total initial resolution elements and
includes dark matter, gas, stars, black holes, and magnetic fields
within a uniform periodic-boundary cube of $51.7$ comoving Mpc per side.
The average mass of the baryonic resolution elements is $8.5 \times 10^4$ $M_{\odot}$,
while the spatial scale of the simulation is set by a gravitational softening 
of the collisionless components (dark matter and stars) 
to be $0.575$ comoving kpc until $z = 1$.
This translates to 144 pc at z = 3, 115 pc at z = 4, 96 pc at z = 5, and 82 pc at z = 6.
As discussed in detail in Appendix B of \citet{Pillepich2019}, the median size of TNG50 galaxies 
represent the resolution-independent outcome of the TNG model, and
the softening per se is not responsible for setting their physical extent, especially at $M_{\star} > 10^8 M_{\odot}$ and $z>1$.
On the other hand, the FWHM of the NIRCam PSF is 521 pc (907 pc) at $z=3$ in the F200W (F356W) band,
470 pc (818 pc) at $z=4$ in the F200W (F356W) band,
424 pc (739 pc) at $z=5$ in the F200W (F356W) band,
and 386 pc (672 pc) at $z=6$ in the F200W (F356W) band.
Thus, our current study is limited in spatial resolution more by observational effects, if anything, than by
the intrinsic resolution of the simulation.

Consistently with TNG50, throughout this work we use cosmological parameters 
consistent with recent Planck measurements \citep[matter density $\Omega_{\rm m} = 0.3089$, 
baryonic density $\Omega_{\rm b} = 0.0486$, cosmological constant $\Omega_{\Lambda} = 0.6911$, 
Hubble constant $H_0 = 100h$ km s$^{-1}$ Mpc$^{-1}$ with $h = 0.6774$, normalization $\sigma_8 = 0.8159$, 
and spectral index $n_{\rm s} = 0.9667$;][]{Planck2016}.

\begin{figure*}[t!]
\centering
\includegraphics[width=\textwidth, trim=0cm 0cm 0cm 0cm , clip=true]{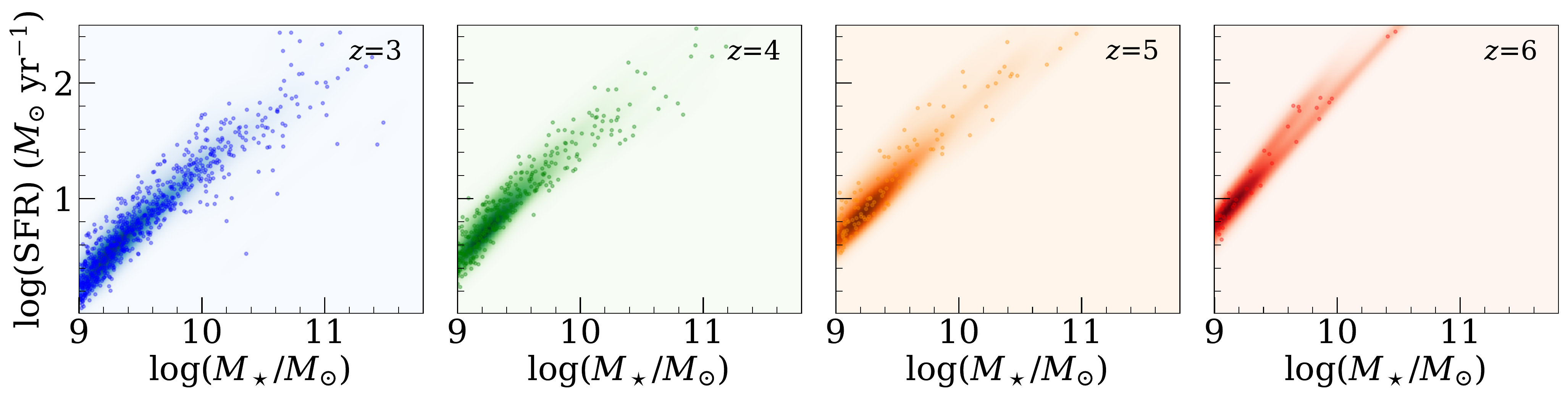}
\caption{Total SFR as a function of the stellar mass for the sample galaxies.
From left to right: $z=3$,  $z=4$,  $z=5$, and  $z=6$.
\label{fig:figure1}}
\end{figure*}

\subsubsection{TNG50 high-$z$ galaxies  \label{sec:section221}}

The TNG50 simulation has already been shown to accurately 
reproduce the morphologies of galaxies at $z<3$ \citep{Tacchella2019, HuertasCompany2019, Varma2022}.
In this work, we consider four TNG50 snapshots that correspond to galaxies at redshift $z=(3,4,5,6)$,
selecting all galaxies with total stellar mass $M_{\star} > 10^9$~$M_{\odot}$ at a given redshift.
The parent sample is composed of 1334 galaxies.
We limit our analysis to galaxies with a total half-mass radius larger than 1~arcsec, with $R_{\rm total} \sim 10 R_{\rm half, \star}$,
which translates to galaxies with a total diameter larger than $\sim64$ pixels when simulated at the angular resolution 
of NIRCam in the F200W filter (see Sect.~\ref{sec:section31}).
With this selection, we exclude from our sample 8\% of galaxies at $z=3$,
5\% of galaxies at $z=4$, 1\% of galaxies at $z=5$, and none of the galaxies at $z=6$.
We checked that no bias is introduced by means of our selection 
in the parameter space of the final sample (i.e., the intrinsic mass-size plane is preserved at every redshift).
In Vega-Ferrero et al.~(\emph{in prep.}), this criterion enables us
to produce cutouts of a fixed size of $64\times64$ pixels for all the galaxies,
which will be ingested into a neural network for deriving galaxy morphologies.

With this selection, the final sample is composed of 1238 galaxies (see Table \ref{tab:table1}). 
To increase the sample statistics, each galaxy is observed in 20 different configurations,
which correspond to 20 random orientations.
In particular, we look at each galaxy under five inclinations $i=(0,45,90,135,180)$ degrees and four azimuths $a=(0, 90,180,270)$ degrees.
Thus, the final sample is composed of $24760$ galaxy projections.
We detail in Table~\ref{tab:table1} their stellar mass median values
and in Fig.~\ref{fig:figure1} we show their total star-formation rate (SFR) distributions as a function of mass at different redshifts,
showing that the vast majority of the sample is composed of star-forming galaxies.


\section{Synthetic images \label{sec:section3}}

In this section, we describe the main steps to produce synthetic images 
tailored for JWST observations starting from cosmological simulations (Fig.~\ref{fig:figure2}).
We first detail the methodology 
for post-processing galaxies in cosmological simulations 
to get noiseless images for NIRCam and MIRI in Sect.~\ref{sec:section31}.
We focus on CEERS observational strategy and 
describe how to create NIRCam-like raw data in Sect.~\ref{sec:section32}.
Finally, we briefly outline the main steps for reducing JWST imaging 
observations in Sect.~\ref{sec:section33}.

\subsection{NIRCam and MIRI noiseless images \label{sec:section31}}

In order to produce noiseless synthetic images of TNG50 high-$z$ galaxies,
we modeled both the light distribution produced by stellar populations and star-forming regions
and the effects of dust on radiation using the most recent version of 
SKIRT Monte-Carlo radiative transfer code \citep[v9.0;][]{Camps2015, Verstocken2017, Camps2020}.

For each galaxy and each snapshot described in Sect.~\ref{sec:section221}, we extract
the corresponding sets of subhalo star particles and gas cells,
while stellar wind particles are ignored.
Each image has a side equal to twice the total half-mass radius of the corresponding galaxy,
but greater than 2~arcsec (see Sect.~\ref{sec:section221}).
The number of pixels is chosen so that the resulting pixel scale matches 
that of JWST instruments. In particular, the spatial resolution is $0.11$~arcsec~px$^{-1}$ for MIRI 
and $0.031$ ($0.063$)~arcsec~px$^{-1}$ for NIRCam short (long) channel.

In the following Sections, we describe the different steps that are required for the radiative transfer calculations.
As a note of caution, several assumptions are necessary to post-process numerical simulations.
We followed standard prescriptions from the literature 
\citep{RodriguezGomez2019, Camps2020, Shen2020, Kapoor2021}, 
but different approaches could be explored \citep[see e.g., ][]{Popping2022}.

\subsubsection{Stellar source}

The primary source of emission is characterized by assigning each stellar particle
a spectral energy distribution (SED) depending on its age in the simulation.
On one side, old star particles ($t > 10$~Myr) are modeled with a SED from the \citet{Bruzual2003}
high-resolution template library and a \citet{Chabrier2003} initial mass function.
In our setup, SKIRT spatially distributes the photons from these sources
using the smoothing length enclosing $32\pm1$ 
nearest stellar particles for all stellar particles within the galaxy.
On the other side, young stellar particles ($t < 10$~Myr) are treated as unresolved regions of the interstellar medium (ISM)
and are modeled with a SED from the \texttt{MAPPINGS~III} library \citep{Groves2008},
which describes emission from both H~II regions and the 
photo-dissociation region surrounding the star-forming core as well as
absorption by gas and dust in the birth clouds of young stars.
The \texttt{MAPPINGS~III} models are parametrized by five parameters: 
(i) the metallicity of the star particle, (ii) the star formation rate (SFR),
(iii) the compactness $C_0$ of the H{\footnotesize II} region, (iv) the pressure $P_0$ of the ISM,
and (v) the covering fraction of the photo-dissociation region $f_{\rm PDR}$.
In this work, we assume the metallicity to be the same as the birth environment 
of the young star particle \citep{Vogelsberger2020}, we
assume that the SFR remains constant during the 10~Myr lifetime 
of the H{\footnotesize II} region \citep{RodriguezGomez2019}, and we used typical values of $\log_{10}\sl{C_0} = 5$,
$\log[(P_0/k_B)/{\rm cm}^{-3} {\rm K}] = 5$, and $f_{\rm PDR} = 0.2$
\citep{Groves2008, Jonsson2010}.

\subsubsection{Dust modeling}

\begin{figure*}[t!]
\centering
\includegraphics[width=\textwidth, trim=0cm 0cm 0cm 0cm , clip=true]{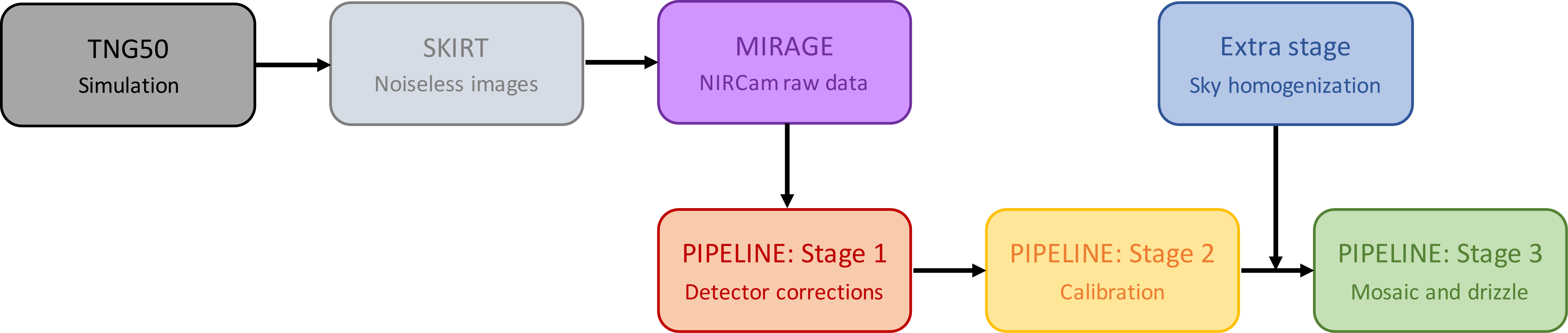}
\caption{Workflow of the proposed forward modeling, from cosmological simulations to reduced NIRCam-like images, which mimic CEERS observations.
\label{fig:figure2}}
\end{figure*}

TNG50 does not include dust physics, hence we model the
distribution of dust in the ISM using the properties (position, density, and metallicity) of the Voronoi gas cells.
We select cold and star-forming gas cells ($T < 8000$~K and SFR$>0$) and derive their metal and mass
distribution \citep{Kapoor2021}. Following \citet{Shen2020}, 
we assume that the metals in the ISM trace the dust component and convert 
the metal mass into the dust mass with a dust-to-metal ratio, which depends 
on redshift as $f_{\rm dust} = 0.9\times(z/2)^{-1.92}$ \citep{Vogelsberger2020}.  
This empirical trend has been calibrated 
based on the observed UV luminosity functions of IllustrisTNG galaxies 
at redshift $z=2-10$.
Thus, we set the dust density to be $\rho_{\rm dust} = f_{\rm dust} Z \rho_{\rm gas}$
for cold and star-forming gas cells and zero for all the other gas cells.
The dust composition is modeled with the dust mix of \citet{Zubko2004},
which includes graphite grains, silicate grains, and polycyclic aromatic 
hydrocarbons \citep{Camps2016,RodriguezGomez2019}.
This choice of multigrain models implies a correction
in the dust density, which was decreased by 25\% as described in \citet{Shen2020}.
Each type of dust has been discretized in 10 bins of grain sizes.
The properties of the dust mix (sizes and abundances) are chosen
so as to reproduce the dust properties of the Milky Way.
Furthermore, for a more realistic description of UV to IR luminosities,
we allow dust grains to be stochastically heated and decoupled
from local thermal equilibrium, and we also include
dust self-absorption and re-emission.
The dust emission and self-absorption are repeated iteratively
until the total luminosity absorbed by dust converges at a $<10\%$ level.

The dust density distribution is then discretized using an octree grid
with a minimum refinement level of 3 and a maximum refinement level of 12 \citep{Kapoor2021}.
We set a maximum cell dust fraction value of $10^{-6}$.
It is worth noticing that within each grid cell, 
every physical quantity (e.g., the dust density, the radiation field, etc.) 
is assumed to be uniform.

\subsubsection{Image signal-to-noise ratio}

The quality of the final synthetic images is directly correlated to the number
of photon packets employed in the radiative transfer simulation.
SKIRT employs a wavelength grid for launching at each position $N_{\rm p}$ photon packets.
The photons are collected by the detector after propagating through the resolved ISM
and randomly interacting with the dust cells.
On the one side, the more photons are employed in the simulation 
the higher the signal-to-noise ratio (SNR) of the final image.
On the other side, $N_{\rm p}$ also drives the simulation run time.
We tweak the number of required photon packets by looking at the 
relative error statistic described in \citet{Camps2020}.
A good compromise between sufficient SNR and an acceptable simulation run time
was found setting $N_{\rm p} = 10^8$ photon packets per galaxy.

\subsubsection{Synthetic data products}

The post-processing of TNG50 galaxies allows us to create a set of synthetic observables 
comprised of noiseless images in multiple bands at different spatial resolutions, 
including spatially integrated fluxes of the total galaxy and each source 
component (i.e., primary and secondary emissions).
In detail, we take advantage of SKIRT flexibility and observe each galaxy from 20 configurations.
It is worth noticing that each additional configuration increases both the simulation run time and the memory consumption. 
At every observer position, we place a detector to mimic both NIRCam and MIRI observations.
Each instrument provides both the spatially integrated flux and the
images for each of NIRCam and MIRI filters (narrow, medium, and broadband) at the 
corresponding pixel scale (see Sect.~\ref{sec:section221}). 
Thus, our TNG50 mock catalogue\footnote{Data publicly released at \url{https://www.tng-project.org/costantin22}.} 
comprises noiseless images of 1238 galaxies at redshift $z=(3,4,5,6)$
seen from 20 different configurations in 27 NIRCam and 9 MIRI bands. 
Supplementary, a super-resolved version ($0.01$~arcsec~px$^{-1}$) of each image is also available.

\subsection{NIRCam raw images \label{sec:section32}}

We mimic NIRCam observations of TNG50 galaxies using the \texttt{mirage}\footnote{\texttt{mirage}
is an open-source Python package developed by STScI and available at \url{https://github.com/spacetelescope/mirage}.}
\citep[Multi Instrument Ramp Generator;][]{Hilbert2019} simulator v2.2.1.
This tool allows us to simulate imaging data that have the instrumental noise effects
that will be present in in-flight data. The final raw data are created by
adding astronomical sources to real dark current data from ground testing.
The scenes could represent different levels of complexity, from point sources and S\'ersic-profile galaxies,
to more realistic morphologies from arbitrary fits images, like in our case.

Briefly, the simulation requires four stages: (1) it creates an input file
with the observational settings exported from the Astronomer’s Proposal Tool (APT);
(2) it creates a seed image, which corresponds to a noiseless image of the scene;
(3) it prepares the dark current exposure; 
(4) it produces the final raw image (\texttt{\_uncal} extension) considering the noise 
contributions both from the background and the detectors. 

Our final scene is composed of a galaxy in the center and three reference stars,
which are used to align the three dithered images in the final mosaic (Sect.~\ref{sec:section334}) and 
build the observed point-spread function (PSF) for drizzled images (Appendix~\ref{sec:appendixB}).

\subsection{Data reduction and calibration \label{sec:section33}}

Data simulated with \texttt{mirage} can be run through the JWST calibration pipeline v1.4.6 (CRDS v11.10.1)\footnote{The 
JWST Data Reduction Pipeline is an open-source Python package developed by STScI
and available at \url{https://jwst-pipeline.readthedocs.io/en/latest/index.html}.}, 
mimicking the data reduction strategy to be used for in-flight data.
The calibration and reduction procedure takes place in three different stages,
which are briefly detailed in the following sections (Fig.~\ref{fig:figure2}).
The step-by-step description provided hereafter concerns the images obtained with detectors B1 and B5, 
which are used for the NIRCam filters F200W and F356W, respectively.

\subsubsection{Stage 1: Detector corrections}

The first stage of the JWST pipeline it is usually referred to as “ramps-to-slopes” processing,
since it reads the non-destructive detector and it translates the integrations containing 
the accumulating counts (ramps) to uncalibrated images (slopes) in units of DN sec$^{-1}$.
All the detector corrections are applied to all exposure types (e.g., imaging, spectroscopic, etc.). 

In stage 1 multiple detector corrections are applied.
The data quality initialization step populates the data quality mask,
allowing the identification of dead pixels in the detector, highly non-linear pixels, and all sorts of quality flags.
The saturation flagging step loops over all integrations within an exposure to
flag all pixels where the signal is above the saturation limit.
In the case of an integration with multiple groups, once one of them is flagged as saturated,
also all subsequent groups for that pixel are flagged.
For each integration in the input science data, the data are corrected group-by-group 
for interpixel capacitance. This step is usually skipped by default, but it was switched on in our reduction.
The superbias subtraction step removes pixel by pixel the detector bias from every group in every integration
of the science ramp data. 
The linearity correction step corrects science data values for detector non-linearity
on a pixel-by-pixel, group-by-group, integration-by-integration basis.
Based on a model, this step computes the number of traps that are expected to have captured or released a charge during an exposure. 
The released charge is proportional to the persistence signal, and this will be subtracted (group by group) from the science data.
The persistence correction step subtracts group-by-group from the science data
the number of traps that are expected to have captured or released a charge during an exposure.
This step is of key importance for real observations, but it could have been skipped 
for our particular case, since \texttt{mirage} does not simulate this effect.
The dark current subtraction step removes dark current from an exposure, group by group.
The cosmic ray flagging step looks for outliers in the ramp of each pixel. 
It compares the signal in all the groups within an integration and it flags those with
deviating more than $5\sigma$.
The ramp fitting step produces a slope image for each integration (\texttt{\_rate} extension).

\begin{figure*}[t!]
\centering
\includegraphics[width=\textwidth, trim=0cm 0cm 0cm 0cm , clip=true]{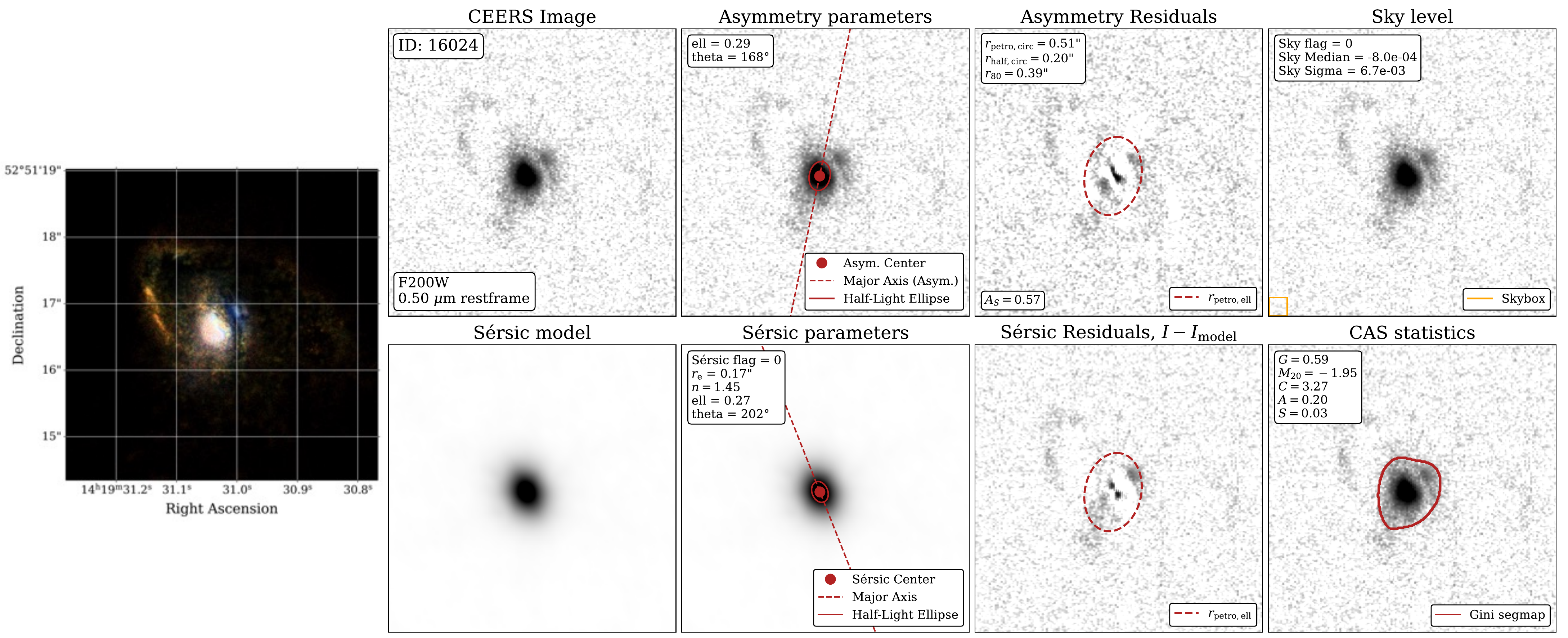}
\caption{Example of morphological parameters derived with \texttt{statmorph} for galaxy ID 16024 ($z=3$, $M_{\star} \sim 10^{11} M_{\odot}$, $i=0$, $a=0$) in the F200W filter
at 0.015 arcsec px$^{-1}$. On the left panel, we show the RGB synthetic image of the galaxy (0.01 arcsec px$^{-1}$), 
created using the F090W, F150W, and F200W noiseless data.
Top row (from left to right): the CEERS image oriented with North up and East left, 
the asymmetry parameters (the galaxy center is shown as a red point, the half-light ellipse is shown as a solid red line, and the orientation
of the major axis is shown as a dashed red line),
the asymmetry residuals and size measurements (elliptical $r_{\rm petro}$ shown as a red dashed line), and sky statistics (median and rms in MJy sr$^{-1}$)
with the sky region (orange box).
Bottom row (from left to right): the noiseless S\'ersic model, the S\'ersic best fit parameters 
(the galaxy center is shown as a red point, the half-light ellipse derived from the S\'ersic model is shown as a solid red line, 
and orientation of the major axis is shown as a dashed red line),
S\'ersic residuals, and non-parametric statistics (CAS, Gini, M$_{20}$) with the Gini segmentation map 
(red contour) derived as described in \citet{RodriguezGomez2019}.
\label{fig:figure3}}
\end{figure*}

\subsubsection{Stage 2: Calibration}

In stage 2, spectroscopic and imaging data are treated differently.
For all JWST instruments, the imaging stage 2 is designed for calibrating individual slope images,
providing data in units of MJy sr$^{-1}$. 
Firstly, the World Coordinate System (WCS) creation step allows to 
transform each position on the detector to a position in a world coordinate frame.
The flat fielding step corrects for pixel-to-pixel sensitivity variations,
dividing the science data set by a flat-field reference image.
The photometric calibration step converts each image from units of countrate (ADU s$^{-1}$) to surface brightness (MJy sr$^{-1}$).
The resample step corrects the flux-calibrated slope images from the effects of instrument distortions.
As a result, individual fully calibrated (but unrectified) exposures are provided (\texttt{\_cal} extension).

\subsubsection{Extra Stage: Sky homogenization}

Before stacking the individual frames obtained at a given sky position, we flattened and removed the background in all calibrated data images. 
This is done by fitting the image to a 3rd order two-dimensional polynomial, 
isolating the background pixels by masking all objects present in the field 
(in our case, only one galaxy and three reference stars). 
The final product of this step is a calibrated image with flat null background. 

\subsubsection{Stage 3: Mosaic \label{sec:section334}}

In stage 3, the calibrated data are combined according to the dither or mosaic pattern into a single 
rectified (distortion-corrected) image. A final step is applied to correct for 
astrometric alignment, background matching, and outlier rejection.
As a result, a resampled and fully calibrated image is provided (\texttt{\_i2d} extension), as well as
a source catalog and a two-dimensional segmentation map.
In this work, we combined the three exposures and created 
two datasets$\footnote{The v1.0 and v1.1 datasets are publicly released at \url{https://www.lucacostantin.com/OMEGA}.}$: 
the first one (v1.0) has images at nominal angular scale (0.063 and 0.031 arcsec px$^{-1}$ for short and long channels, respectively),
while for the second one (v1.1) 
the images were drizzled to have a higher angular resolution (half-nominal),
mimicking the strategy foreseen for CEERS observations (0.030 and 0.015 arcsec px$^{-1}$ for short and long channels, respectively). 


\section{Results and discussion \label{sec:section4}}

In this section we describe the main results of our analysis, focusing 
on the potential of CEERS in characterizing the morphology of galaxies
at different redshifts ($3 \leq z \leq 6$) and wavelengths (F200W, F356W).
We discuss the measurements of galaxy sizes and morphological
properties at half-nominal angular resolution,
also investigating the effect that drizzling NIRCam images has 
on our measurements (see Appendix~\ref{sec:appendixA}).

\begin{figure*}[t!]
\centering
\includegraphics[width=\textwidth, trim=0cm 0cm 0cm 0cm , clip=true]{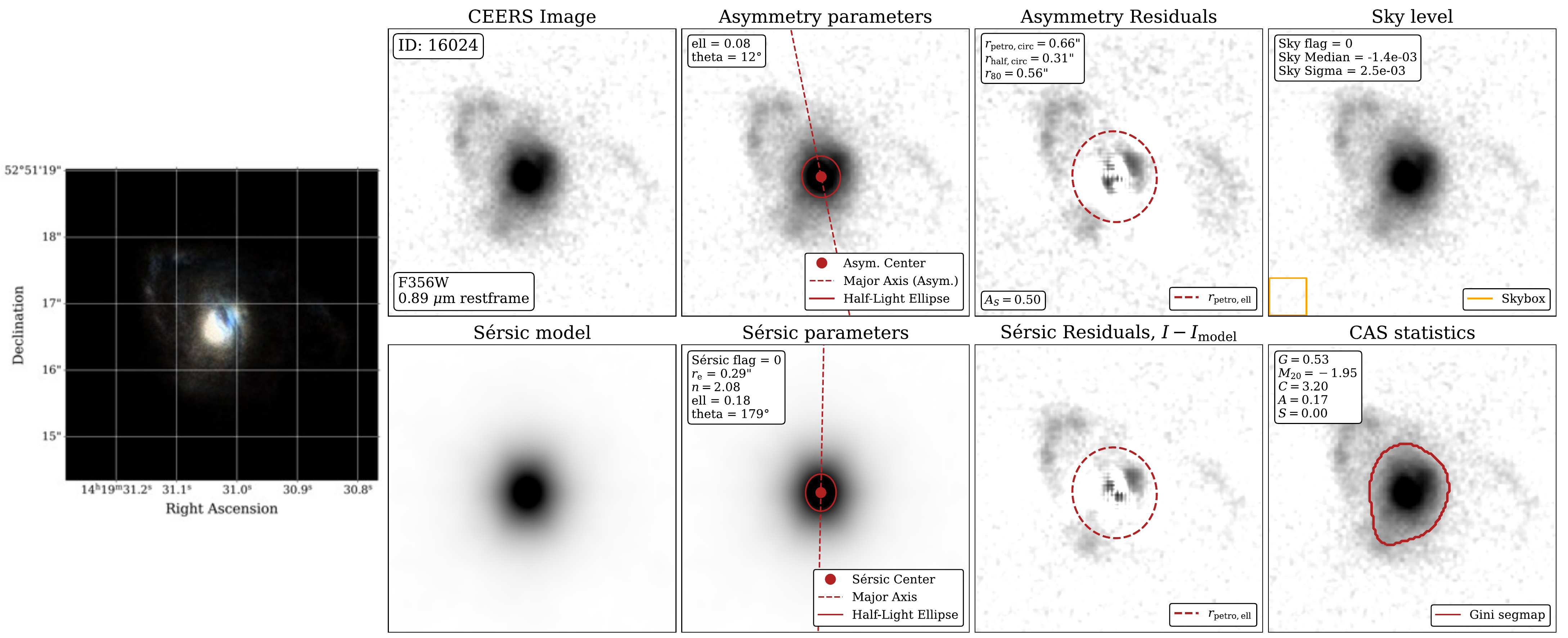}
\caption{As in Fig.~\ref{fig:figure3}, but for the F356W filter at 0.030 arcsec px$^{-1}$.
The RGB image is obtained using the F277W, F356W, and F444W filters.
\label{fig:figure4}}
\end{figure*}

\subsection{Morphological measurements \label{sec:section41}}

We derive the parametric and non-parametric morphology of our sample galaxies
using the standard configuration of \texttt{statmorph}\footnote{\texttt{statmorph} is available at \url{https://statmorph.readthedocs.io}.}, 
a Python package developed by \citet{RodriguezGomez2019}
and optimized to compute the optical morphologies of galaxies.
All the details about the different sets of parameters derived with \texttt{statmorph} as
well as the discussion about their definitions are fully explained by \citet{RodriguezGomez2019}.

Regarding the data quality, \texttt{statmorph} flags ``bad measurements'' either if there is a problem with the basic measurements (e.g., artifacts in the image
or bad sky subtraction) or if there is a problem during the S\'ersic fit, which could happen for galaxies with
very irregular morphologies. Furthermore, we visually flag galaxies that are poorly detected 
or which have highly irregular segmentation maps. 
In the following, we will consider only galaxies
with no flags, taking into account their redshift, filter, and angular resolution.

We recall that the primary objective of this work is to provide mock-observed
morphological parameters, of simulated galaxies, for an easier comparison with the first JWST observations. 
We focus on different size estimators (e.g., the Petrosian radius $r_{\rm petro}$, the half-light radius from aperture photometry 
$r_{\rm half, phot}$, and the effective radius from the S\'ersic fit $r_{\rm e}$), but we also derived the Gini coefficient,
the M$_{20}$ statistic, the F(G,$M_{20}$) statistic, the galaxy concentration $C$, 
asymmetry $A$, smoothness $S$, and the S\'ersic index $n$ (see Appendix~\ref{sec:appendixA}).
For each of these measurements, each galaxy is considered a single-component system.
As a complementary product of this work, we released the morphological catalog with all measured 
parameters$\footnote{Data publicly released at \url{https://www.lucacostantin.com/OMEGA}.}$.

In Figures \ref{fig:figure3} and \ref{fig:figure4} we provide an example of the
measured parameters for the galaxy ID 16024 ($z=3$, $M_{\star} \sim 10^{11}~M_{\odot}$, $i=0$, $a=0$) 
in the F200W and F356W bands, respectively. Focusing on the RGB synthetic images, 
it can be appreciated that the galaxy shows a prominent tidal feature on the west side
which is fading at longer wavelengths. This structure is detected 
both in the F200W and F356W bands, even though it is quite faint.
Moreover, a prominent dust lane is situated on the east side of the galaxy, 
visible at short and long wavelengths both in the synthetic and calibrated images.
The galaxy can be modeled with a S\'ersic law with $n\sim1.5$ ($n\sim2$) and $r_{\rm e} \sim 0.15$~arcsec
($r_{\rm e} \sim 0.3$~arcsec) in the F200W (F356W) band.
The intrinsic half-mass radius ($R_{\rm half, \star}$, see Sect.~\ref{sec:sec_42}) is 0.79~kpc,
which led to $r_{\rm e}$/$R_{\rm half, \star}$ = 1.75.
From the residual maps, we can infer that the galaxy is probably hosting a
central bulge embedded in an extended disk.
Moreover, the galaxy is supported by rotation ($v_{\rm max}/\sigma \sim 2$),
consistent with a late-type system of disk-like morphology, but already showing a prominent bulge.

\begin{deluxetable*}{cDDDDD}
\tablecaption{Median sizes of galaxies measured in the F200W band at different redshift. 
\label{tab:table2}}
\tablehead{
\colhead{$z$} & \multicolumn2c{$R_{\rm half, \star}$} &\multicolumn2c{$r_{\rm e}$} & \multicolumn2c{$r_{\rm petro}$} & \multicolumn2c{$r_{\rm e}$/$R_{\rm half, \star}$} & \multicolumn2c{$r_{\rm petro}$/$r_{\rm e}$} \\
\colhead{} & \multicolumn2c{(kpc)} & \multicolumn2c{(kpc)} & \multicolumn2c{(kpc)} & \multicolumn2c{} & \multicolumn2c{} 
}
\decimalcolnumbers
\phd
\startdata
3 	 	&	$1.79^{+1.12}_{-0.72}$	&	$1.71^{+0.80}_{-0.50}$	&	$4.01^{+1.58}_{-0.98}$		&	1.00^{+0.53}_{-0.36}		&	$2.32^{+1.43}_{-0.86}$	   	\\
4 		&	$1.44^{+0.83}_{-0.68}$	& 	$1.29^{+0.59}_{-0.39}$	&	$3.09^{+1.17}_{-0.77}$		&	0.97^{+0.50}_{-0.35}		&	$2.27^{+1.48}_{-0.81}$	   	\\
5 		& 	$1.27^{+1.03}_{-0.63}$	& 	$1.09^{+0.50}_{-0.36}$	&	$2.60^{+1.04}_{-0.75}$	   	&	0.92^{+0.57}_{-0.38}		&	$2.15^{+1.81}_{-0.90}$	   	\\
6 		&	$1.14^{+0.45}_{-0.90}$	& 	$0.80^{+0.36}_{-0.38}$	&	$1.94^{+0.94}_{-0.58}$	   	&	0.87^{+0.99}_{-0.31}		&	$2.08^{+4.08}_{-0.78}$	   	\\
\enddata
\tablecomments{
(1) Redshift. 
(2) Median value and 16th-84th percentile range of half-mass radius.
(3) Median value of effective radius.
(4) Median value of elliptical Petrosian radius.
(5) Median value of the ratio $r_{\rm e}/R_{\rm half, \star}$.
(6) Median value of the ratio $r_{\rm petro} / r_{\rm e}$.
}
\end{deluxetable*}

\begin{deluxetable*}{cDDDDD}
\tablecaption{Median sizes of galaxies measured in the F356W band at different redshift. Columns as in Table~\ref{tab:table2}.
\label{tab:table3}}
\tablehead{
\colhead{$z$} & \multicolumn2c{$R_{\rm half, \star}$} &\multicolumn2c{$r_{\rm e}$} & \multicolumn2c{$r_{\rm petro}$} & \multicolumn2c{$r_{\rm e}$/$R_{\rm half, \star}$} & \multicolumn2c{$r_{\rm petro}$/$r_{\rm e}$} \\
\colhead{} & \multicolumn2c{(kpc)} & \multicolumn2c{(kpc)} & \multicolumn2c{(kpc)} & \multicolumn2c{} & \multicolumn2c{} 
}
\decimalcolnumbers
\phd
\startdata
3 	 	&	$1.84^{+1.23}_{-0.76}$	&	$1.74^{+0.79}_{-0.52}$		&	$4.32^{+1.63}_{-1.02}$	&	0.96^{+0.51}_{-0.32}		&	$2.36^{+1.36}_{-0.73}$	   	\\
4 		&	$1.53^{+0.97}_{-0.69}$	& 	$1.35^{+0.64}_{-0.41}$		&	$3.40^{+1.23}_{-0.77}$	&	0.92^{+0.44}_{-0.32}		&	$2.26^{+1.37}_{-0.73}$	   	\\
5 		&	$1.42^{+1.47}_{-0.74}$	& 	$1.15^{+0.57}_{-0.35}$		&	$2.93^{+1.14}_{-0.73}$	&	0.84^{+0.46}_{-0.34}		&	$2.15^{+1.56}_{-0.91}$	   	\\
6 		&	$1.39^{+0.69}_{-0.86}$	& 	$0.87^{+0.44}_{-0.33}$		&	$2.35^{+0.87}_{-0.62}$	&	0.75^{+0.56}_{-0.28}		&	$1.93^{+3.85}_{-0.69}$	   	\\
\enddata
\end{deluxetable*}

\subsection{Size definitions \label{sec:sec_42}}

In this work, we discuss the evolution of sizes in high-$z$ massive galaxies in TNG50 and
consider the following size measurements: the effective radius from the S\'ersic fit ($r_{\rm e}$),
the Petrosian radius measured on elliptical apertures ($r_{\rm petro}$), and the half-light radius derived from
elliptical aperture photometry ($r_{\rm half, phot}$; see Appendix~\ref{sec:appendixB}). 
Finally, we used the three-dimensional stellar half-mass radius
\citep[$R_{\rm half, \star}$;][]{Pillepich2019}, which is measured as the three-dimensional radius containing half of the stellar mass 
of all constituent stars gravitationally bound to a galaxy, as a reference for the mass distribution of TNG50 galaxies.

The use of multiple size definitions in the literature \citep{Sersic1968, Petrosian1976, Ferguson2004, Bouwens2004, Graham2005, Chamba2020}
reflects the ambiguity in defining the total extent of a galaxy, which is assumed to contain its entire flux.
For high-$z$ galaxies, the task of measuring their sizes is even more challenging, 
because (1) the irregular/chaotic morphology makes it difficult to define the galaxy center and perform aperture photometry;
(2) the outskirt of the galaxy has very low surface-brightness features, usually 
not detected;
(3) aperture photometry does not take into account the effect of the PSF (see Appendix~\ref{sec:appendixB}),
while one-component (or multi-component) fitting is difficult because of the complex morphology of these galaxies;
(4) the redshift of a given galaxy influences the surface density limits \citep[i.e., cosmological dimming;][]{Giavalisco1996, Ribeiro2016},
so it is not straightforward to compare galaxies in a wide range of redshifts.

\begin{figure*}[t!]
\centering
\includegraphics[width=0.8\textwidth, trim=0cm 0cm 0cm 0cm , clip=true]{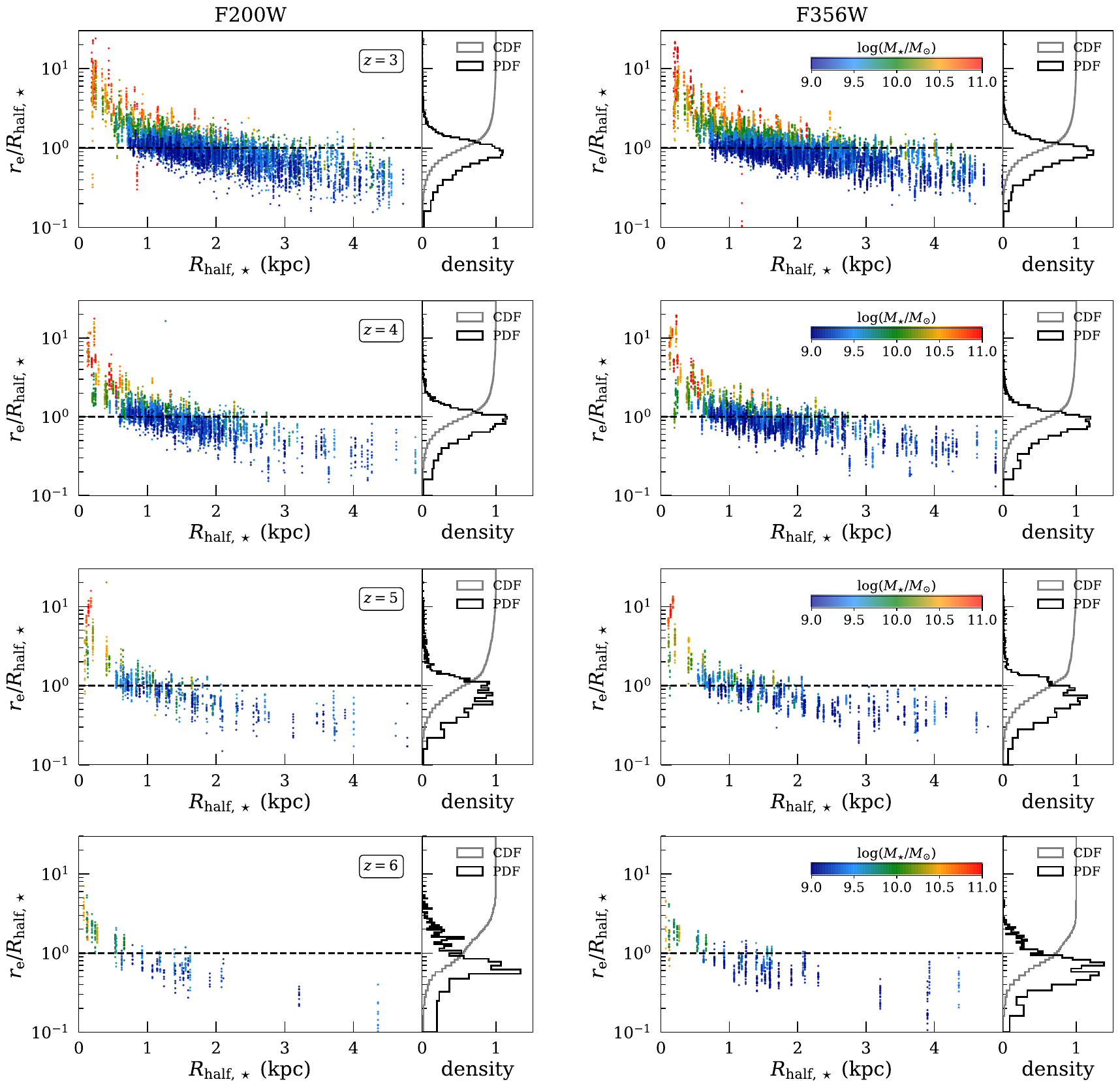}
\caption{Comparison between $R_{\rm half, \star}$ and $r_{\rm e}$ for galaxies at different redshift (from top to bottom)
in the F200W band (left panels) and F356W band (right panels).
Main panels: Data points are color-coded according to the value of the galaxies' total stellar mass.
The black dashed line marks the 1:1 relation of the ratio $r_{\rm e}$/$R_{\rm half, \star}$.
Side panels: probability density function (black solid line) and cumulative density function (gray solid line) of $r_{\rm e}$/$R_{\rm half, \star}$.
\label{fig:figure5}}
\end{figure*}

\begin{figure*}[t!]
\centering
\includegraphics[width=0.7\textwidth, trim=0cm 0cm 0cm 0cm , clip=true]{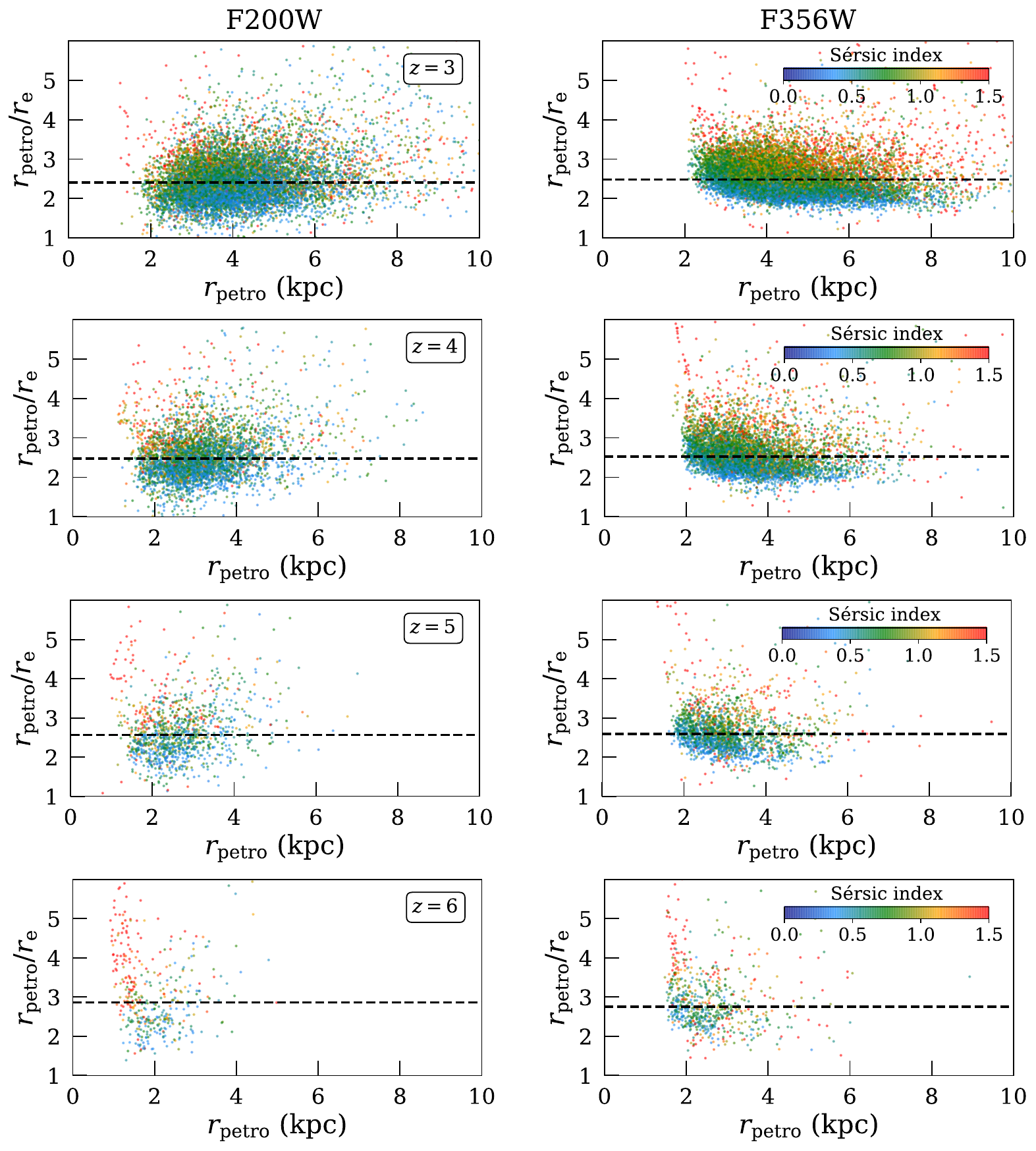}
\caption{Comparison between $r_{\rm petro}$ and $r_{\rm e}$ for galaxies at different redshift (from top to bottom)
in the F200W band (left panels) and F356W band (right panels).
Data points are color-coded according to the value of the S\'ersic index.
The black dashed line marks the median value of the ratio $r_{\rm petro}$/$r_{\rm e}$.
\label{fig:figure6}}
\end{figure*}

\begin{figure*}[t!]
\centering
\includegraphics[width=0.7\textwidth, trim=0cm 0cm 0cm 0cm , clip=true]{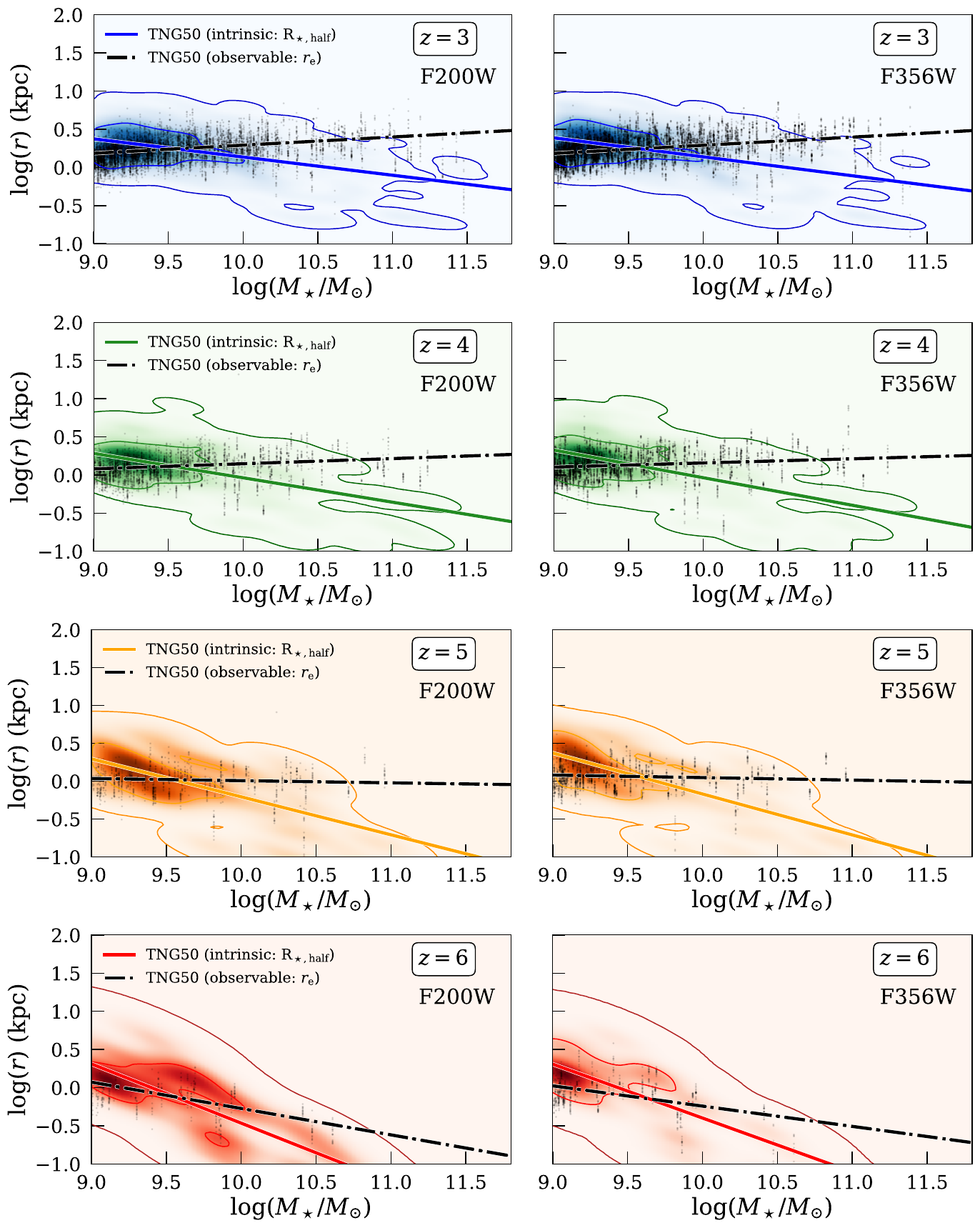}
\caption{Mass-size relations from redshift $z=3$ (top panels) to $z=6$ (bottom panels).
The left panels stand for sizes ($r_{\rm e}$ and R$_{\rm \star, half}$) measured in the F200W band and
the right panels stand for those measured in the F356W band. The blue, green, orange, and red
distributions and contours represent the values of R$_{\rm \star, half}$ from TNG50,
while the black points stand for the values of $r_{\rm e}$ measured on fully calibrated images.
In each panel, the best-fit relation from Eq.~\ref{eq:1} is shown with a colored solid line,
while the one obtained from the mock observed TNG50 galaxies from Eq.~\ref{eq:2} is shown as a dashed black line.
\label{fig:figure7}}
\end{figure*}

\begin{deluxetable}{ccccc}
\tablecaption{Coefficients of the linear regression obtained for the mass-size relation at different redshifts in the F200W band. 
\label{tab:table4}}
\tablehead{
\colhead{$z$} & \colhead{$a_0$} & \colhead{$b_0$} & \colhead{$a_{\rm obs}$} & \colhead{$b_{\rm obs}$} \\
\colhead{} & \colhead{(kpc $M_{\odot}^{-1}$)} & \colhead{(kpc)} & \colhead{(kpc $M_{\odot}^{-1}$)} & \colhead{(kpc)} 
}
\decimalcolnumbers
\phd
\startdata
3 	 	&		$-0.24$		&	2.48		&	0.09			&	$-0.65$		\\
4 	 	&		$-0.33$		&	3.27		&	0.05			&	$-0.39$	   	\\
5 	 	&		$-0.55$		&	5.25		&	$-0.05$		&   	0.44			\\
6 	 	&		$-0.78$		&	7.29		&	$-0.27$		&	2.44	   		\\
\enddata
\tablecomments{
(1) Redshift. 
(2) Slope of the linear trend from Eq.~\ref{eq:1}.
(3) Intercept of the linear trend from Eq.~\ref{eq:1}.
(4) Slope of the linear trend from Eq.~\ref{eq:2}.
(5) Intercept of the linear trend from Eq.~\ref{eq:2}.
}
\end{deluxetable}

\begin{deluxetable}{ccccc}
\tablecaption{As in Table~\ref{tab:table4}, but for the F356W band. 
\label{tab:table5}}
\tablehead{
\colhead{$z$} &  \colhead{$a_0$} & \colhead{$b_0$} & \colhead{$a_{\rm obs}$} & \colhead{$b_{\rm obs}$} \\
\colhead{} & \colhead{(kpc $M_{\odot}^{-1}$)} & \colhead{(kpc)} & \colhead{(kpc $M_{\odot}^{-1}$)} & \colhead{(kpc)} 
}
\decimalcolnumbers
\phd
\startdata
3 	 	&		$-0.26$		&	2.72		&	0.11			&	$-0.80$	   	\\
4 	 	&		$-0.35$		&	3.48		&	0.06			&	$-0.47$	   	\\
5 	 	&		$-0.55$		&	5.27		&	$-0.03$		&	0.34	   		\\
6 	 	&		$-0.71$		&	6.71		&	$-0.27$		&	2.42	   		\\
\enddata
\end{deluxetable}

\subsection{Intrinsic and observable sizes}

In this section we characterize the sizes of our galaxies by means of the Petrosian radius and the effective radius,
describing the goodness/reliability of our measurements and the interplay between $r_{\rm petro}$ and $r_{\rm e}$.
Throughout this section, we will consider sizes measured on images at half-nominal angular resolution.
At each redshift, we report the median values of $r_{\rm e}$, $r_{\rm petro}$, and $R_{\rm half, \star}$ 
in Tables~\ref{tab:table2} and \ref{tab:table3} for the F200W and F356W bands, respectively.

In Fig.~\ref{fig:figure5}, we present the evolution of $r_{\rm e}$ as a function of redshift and wavelength,
comparing its distribution with the one of $R_{\rm half, \star}$.
We find that the ratio $r_{\rm e}$/$R_{\rm half, \star} \sim 1$ at $z=3$ and $z=4$
(see also Figs.~\ref{fig:figure9} and \ref{fig:figure10}), and it starts to deviate at higher redshifts.
The larger deviations are for small galaxies ($R_{\rm half, \star} < 1$~kpc) and for large ones ($R_{\rm half, \star} > 3$~kpc).
In particular, the size of the most massive galaxies (e.g., $\log(M_{\star}/M_{\odot}) > 10.5$ at $z=3$) are overestimated,
with deviations as large as  $r_{\rm e}$/$R_{\rm half, \star} \sim 4-5$ \citep[see also ][]{Wu2020}. 
This behavior is seen both in the F200W and in the F356W band.
In these galaxies, the light distribution is poorly fit by a single component, which 
overestimates the galaxy's size and underestimates the central light concentration (see Appendix~\ref{sec:appendixC}).
This could be interpreted as an indication of the build-up of the first (compact) bulges in disk galaxies,
which happens at earlier cosmic time in more massive galaxies \citep{Tacchella2015, Costantin2021, Costantin2022}.
Besides these caveats, the trend presented in Fig.~\ref{fig:figure5} shows that $r_{\rm e}$ is a good diagnostic for a galaxy's mass-weighted size.

Regarding the effect that drizzling NIRCam images could have in retrieving their structural parameters,
we find that a galaxy's size increases by a factor up to 25\% if measured 
on images at nominal angular resolution as compared to that measured on images at half-nominal angular resolution (see Appendix~\ref{sec:appendixA}).
But, the size distribution at half-nominal angular resolution matches the intrinsic one retrieved from TNG50
(Fig.~\ref{fig:figure5} and Figs.~\ref{fig:figure9}-\ref{fig:figure10}).
Thus, we propose to discuss the structural evolution of our galaxies using the measured morphology on the drizzled images.

Furthermore, we quantify the difference between the sizes derived
from $r_{\rm e}$ and $r_{\rm half, phot}$ (see Appendix~\ref{sec:appendixB}). 
In particular, we find that $r_{\rm e}$/$r_{\rm half, phot} < 1$
mostly at all radii, presenting the largest deviations at small radii (up to 80-90\%).
Indeed, without considering any PSF correction,
$r_{\rm half, phot}$ is not probing galaxy sizes smaller than $\sim1$~kpc.

In Fig.~\ref{fig:figure6} we illustrate the interplay between $r_{\rm petro}$ and $r_{\rm e}$ 
in describing the sizes of our galaxies. We quantify that, independent of redshift and photometric band, 
on average $r_{\rm petro} > 2 \times r_{\rm e}$ (see Tables~\ref{tab:table2} and \ref{tab:table3}).
This trend, and the relation between $r_{\rm e}$ and $R_{\rm half, \star}$, raises a note of caution in limiting
spectroscopic studies in the region covered by the effective radius,
since this region is not even probing half of the mass distribution of the galaxy,
or defining the total extension of the galaxy with a totally arbitrary factor (e.g., $2r_{\rm e}$).
Indeed, the extension of the galaxy depends not only on the effective radius, 
but also on the shape of the surface-brightness profile \citep[i.e., the light and mass distributions;][]{Trujillo2020}.
As shown in Fig.~\ref{fig:figure6}, the ratio $r_{\rm petro} / r_{\rm e}$ depends on the value of the S\'ersic index,
since galaxies with $n<1$ show $r_{\rm petro} / r_{\rm e} < 2.5$,
while galaxies with $n>1$ show $r_{\rm petro} / r_{\rm e} > 2.5$. For example, in the F200W band at $z=3$, the median 
values of these ratios are $2.31^{+0.57}_{-0.41}$ ($n<1$) and $2.67^{+0.61}_{-0.46}$ ($n>1$).

Finally, we compare the values of $r_{\rm petro}$ with the ones derived in \citet{Whitney2019} 
for a sample of 49,000 galaxies from the Cosmic Assembly Near-infrared Deep Extragalactic 
Survey \citep[CANDELS;][]{Grogin2011, Koekemoer2011}.
In particular, we see that the median size of our galaxies follows (within $1\sigma$) the redshift evolution 
expected for high-$z$ galaxies in the form of $r_{\rm petro} = 12.62(1+z)^{-0.82}$,
confirming the goodness of our predictions and already available observations 
targeting the ultraviolet rest-frame morphology \citep[see Fig.~9 in][]{Whitney2019}.
This could point to the fact that no (or mild) wavelength evolution
of sizes is expected at these redshifts, at least comparing the restframe ultraviolet and optical morphology.

\begin{figure}[t!]
\centering
\includegraphics[width=0.45\textwidth, trim=0cm 0cm 0cm 0cm , clip=true]{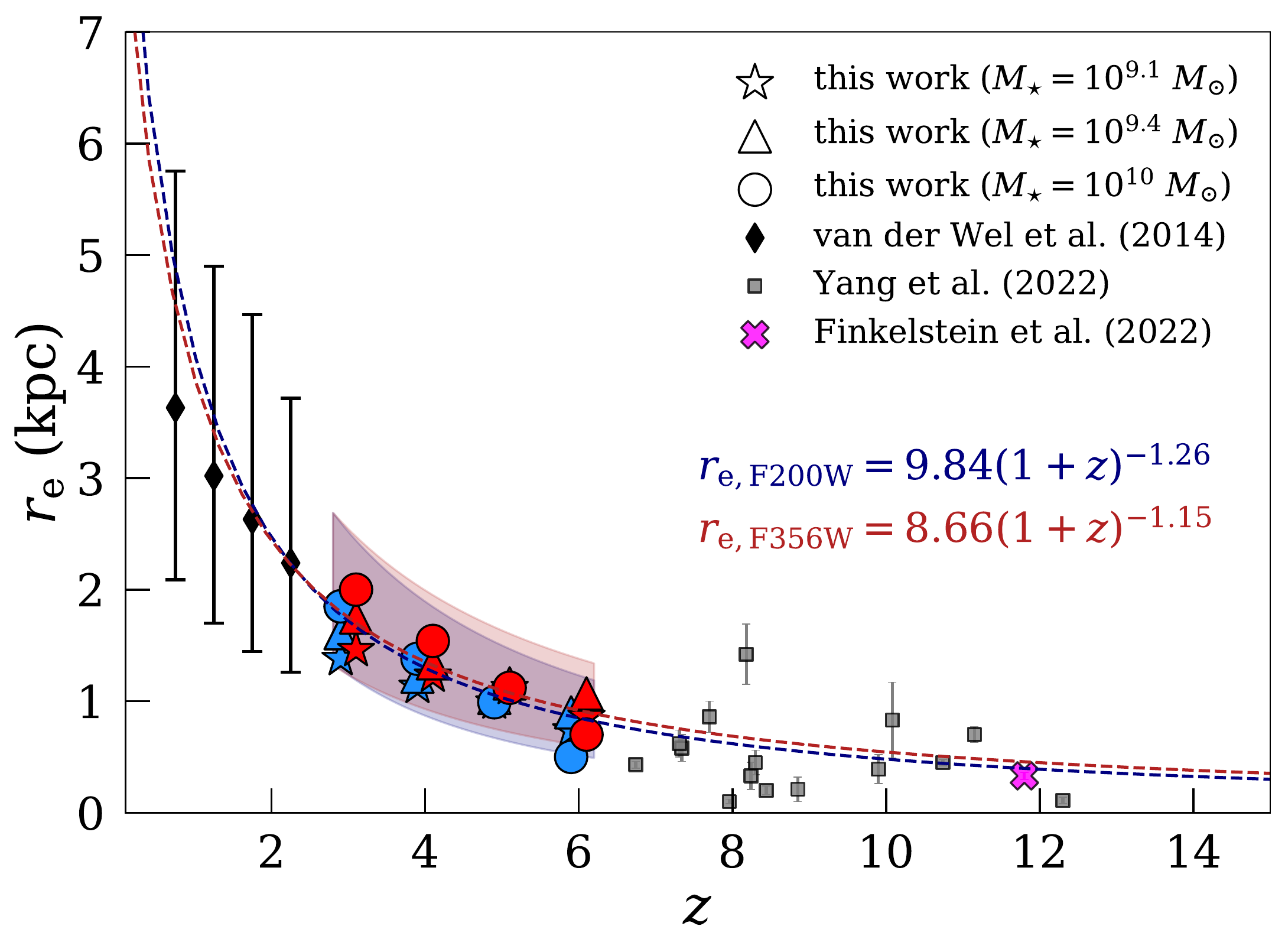}
\caption{Evolution of size ($r_{\rm e}$) through redshift for galaxies in different mass bins.
Stars stand for galaxies with $9 \leq \log(M_{\star}/M_{\odot}) \leq 9.25$,
triangles stand for galaxies with $9.25 < \log(M_{\star}/M_{\odot}) \leq 9.6$,
and dots stand for galaxies with $\log(M_{\star}/M_{\odot}) > 9.6$.
Blue symbols correspond to median values in the F200W band,
while red symbols represent the median value in the F356W band.
The colored shaded regions mark the 16-84th percentile ranges.
Black diamonds correspond to median sizes of late-type galaxies at $z<3$ with 
$\log(M_{\star}/M_{\odot}) \sim 9.75$ \citep{vanderWel2014}.
Gray squares stand for sizes measured in individual galaxies (NIRCam F356W band) at $z>7$ \citep{Yang2022},
while the purple cross corresponds to the Maisie's Galaxy ($z=11.8$) identified in \citet{Finkelstein2022}.
\label{fig:figure8}}
\end{figure}

\subsection{Expected size evolution of high-$z$ galaxies \label{sec:section44}}

Given that the effective radius measured on images at half-nominal angular resolution
is a good proxy for the size of our galaxies as it reproduces the distribution of the half-mass radius,
we provide in this section some ready-to-use diagnostics to discuss the structural evolution of 
high-$z$ galaxies observed with JWST. However, it is worth taking into account that there are (small) systematic deviations
for the very large (and less massive) galaxies and for very compact (and most massive) galaxies.

In Fig.~\ref{fig:figure7}, we describe the redshift evolution of the mass-size relation of high-$z$ galaxies. 
We use the total stellar mass of the galaxy derived from TNG50 and the measured effective radius as a proxy for the galaxies' size.
We show the trend of the half-mass radius (density maps and contours of different colors) and the trend of the observable sizes,
as measured by the effective radius (black symbols).
We derived the coefficients of the best-fit (Tables~\ref{tab:table4} and \ref{tab:table5})
retrieved from parametrizing as a linear regression the intrinsic mass-size relation
\begin{equation}
\label{eq:1}
\log(R_{\rm half, \star}) = a_0 \log(M_{\star}) + b_0 \, ,
\end{equation}
and the observable mass-size relation of the simulated galaxies
\begin{equation}
\label{eq:2}
\log(r_{\rm e}) = a_{\rm obs} \log(M_{\star}) + b_{\rm obs} \, .
\end{equation}
The first result is that, at all redshifts, there is almost no difference in the coefficients describing the 
linear parametrization of the mass-size relation derived at 2~$\mu$m and at 3.56~$\mu$m.
This is true both for the intrinsic and observable parametrization, but with a discrepancy between the two.
Indeed, the intrinsic mass-size relation shows a negative slope at all considered redshifts \citep[see also ][]{Genel2018, Pillepich2019},
while the size-mass trend based on observable $r_{\rm e}$
shows a transition from negative to positive slopes between $z=5$ and $z=4$.
At each redshift, $r_{\rm e}$ is probing the trend of $R_{\rm half, \star}$ up to $\sim 5 \times 10^{10}$ solar masses.
For the more massive galaxies (i.e., the smaller ones in TNG50), $r_{\rm e}$ is overestimating their size (see also Fig.~\ref{fig:figure5}).

The negative slope of the intrinsic luminosity-size relation, apparently counter-intuitive, 
has also been presented in \citet{Marshall2022} for the BLUETIDES simulation \citep{Feng2016}
and in \citet{Roper2022} for the FLARES simulation \citep{Lovell2021}.
At $z\gg3$, according to multiple theoretical and observational predictions 
\citep[e.g., ][]{Dekel2014, Zolotov2015, Tacchella2015, Tacchella2016, Costantin2021},
extreme events of gas compaction in massive galaxies could be responsible for building very massive and small galaxies.
However, this intrinsic trend is in conflict with observations at these redshifts 
\citep[e.g., ][but see also \cite{Mosleh2020}]{Grazian2012, Shibuya2015, Bouwens2022}.
This mismatch could be due to several causes. First, the complex morphology of massive high-$z$ galaxies 
makes it difficult to reproduce the mass distribution with a single-component model. Second, some of
the massive and compact bulges can be obscured by dust (e.g., Figs.~\ref{fig:figure3} and \ref{fig:figure4}).
For instance, \citet{Roper2022} reported that galaxies at $z>5$ can even appear $\sim$50 times 
larger when including dust attenuation.
Finally, it has to be considered that the effective radius is tracing the two-dimensional projection of the 
galaxy light and not its three-dimensional extension.
While we find very mild evolution with wavelength (but targeting the UV/optical rest-frame spectral range),
future JWST observations (e.g., MIRI/JWST), and analyses similar to the one proposed in this work, but targeting the reddest portion of the
electromagnetic spectrum \citep[e.g., ][]{Popping2022}, could help in solving this apparent tension between predictions from cosmological simulations 
and observed sizes of high-$z$ galaxies.

We conclude our discussion by summarizing the size evolution with respect to mass, redshift, and wavelength in Fig.~\ref{fig:figure8}.
Regarding the wavelength trend, at each redshift and mass, sizes measured in the F200W band are consistent (within $1\sigma$) with those measured in the F356W band.
At $z=3$ and $z=4$, this could be explained by the fact that both bands are probing the optical restframe morphology.
At higher redshift, this could be the consequence of the low number of galaxies and the fact that galaxies 
could present (similar) young stellar populations, which dominate the light in both bands.
Regarding the mass trend, at redshift $z=3$ and $z=4$, more massive galaxies are larger than lower-mass ones (see also Fig.~\ref{fig:figure7}).
At $z=5$ galaxies have similar sizes at all masses, while at $z=6$ the low-number statistics from the small TNG50
volume makes it difficult to provide any concluding result.
But, for the most massive galaxies, we see a smooth transition from compact systems at higher redshift to more extended ones at lower redshift.
This trend in the observable size-mass relation of the simulated galaxies, and the mismatch between the mass and light distribution,
could be the consequence that these galaxies have complex morphology, not well described with just one component.
They could present a central bulge, where most of the mass resides, and extended disky structures, biasing the light-weighted galaxy's size
to larger values than the mass-weighted one. Additionally, the central region of these massive galaxies could be obscured by dust \citep{Wu2020}.

Based on this interpretation, we speculate
that from $z=5$ to $z=3$ there could be a turning point in galaxy evolution 
where the physical processes responsible to assemble the first compact systems,
which are supposed to evolve into local elliptical or central bulges, 
are sub-dominant with respect to the physics regulating the disk growth \citep[see e.g., ][]{Costantin2020, Costantin2022}.
Finally, at all masses, galaxies at higher redshift are smaller (more concentrated) than those at lower redshift.
The size evolution of galaxies can be parametrized as $r_{\rm e} = 9.84(1+z)^{-1.26}$ in the F200W band 
and $r_{\rm e} = 8.66(1+z)^{-1.15}$ in the F356W band.
A similar trend is also confirmed by the evolution of the Petrosian radius,
with $r_{\rm petro} = 21.48(1+z)^{-1.21}$ in the F200W band and $r_{\rm petro} = 18.30(1+z)^{-1.04}$ in the F356W band.
The extrapolation of these trends towards lower redshifts is consistent with the 
observational measurements for massive late-type galaxies \citep{vanderWel2014}.
Furthermore, we compared the first measurements of galaxies' sizes in the primordial Universe,
from $z\gtrsim7$ \citep{Yang2022} up to $z\sim12$ \citep{Finkelstein2022}, 
with our estimated trends, finding overall very good agreement.


\section{Summary and Conclusions \label{sec:section5}}

In this work, we created mock observations tailored for the forthcoming JWST observations of high-$z$ galaxies
using a state-of-the-art cosmological simulation (i.e., TNG50). The main goal is to provide predictions for observable
morphological parameters (i.e., by forward modeling stellar light),
focusing on the galaxy's size, to be compared right away with the first JWST datasets.

We generated $\sim 25,000$ synthetic images of $M_{\star} \ge 10^9$~$M_{\odot}$ galaxies from 
the TNG50 cosmological simulation at $z=3$, $z=4$, $z=5$, and $z=6$.
The noiseless images were generated with the radiative transfer code SKIRT v9.0, 
including the effects of dust attenuation and scattering.
The images are available in all the filters of the NIRCam and MIRI instruments of the JWST,
at the corresponding angular resolution, i.e., 0.031 (0.063) arcsec px$^{-1}$ for NIRCam short (long) channel
and 0.11 arcsec px$^{-1}$ for MIRI. A super-resolved version (0.01 arcsec px$^{-1}$) is also available.

As a second step, we simulated mock NIRCam observations following the observational strategy (e.g., noise, dithering pattern, etc.) of CEERS.
Indeed, we processed the synthetic images with \texttt{mirage} and then calibrated them
using the official JWST reduction pipeline, obtaining a set of mock observations in the F200W and F356W bands.
The fully-calibrated images are available both at nominal angular resolution (0.031 arcsec px$^{-1}$  for F200W and 0.063 arcsec px$^{-1}$ for F356W), 
and at half-nominal angular resolution (0.015 arcsec px$^{-1}$  for F200W and 0.030 arcsec px$^{-1}$ for F356W), 
since we also drizzled them mimicking the strategy foreseen for CEERS (and many other wide-field surveys).

Finally, we measured with \texttt{statmorph} different size estimators (i.e., the Petrosian radius and the effective radius from the S\'ersic fit), 
but we also derived the Gini coefficient, the M$_{20}$ statistic, the F(G,M$_{20}$) statistic, the galaxy concentration, asymmetry, 
smoothness, and the S\'ersic index, which we released as a catalog associated with this work.

The expectations of the size evolution of massive galaxies at $3 \le z \le 6$ from the TNG50 simulation of the IllustrisTNG suite
can be summarized in the following points:

\begin{itemize}
\item We found that the (S\'ersic) effective radius is a good proxy for the (three-dimensional) half-mass radius derived from TNG50 at all redshifts.
\item On average, the Petrosian radius is found to be $r_{\rm petro} > 2 \times r_{\rm e}$.
\item The sizes of high-$z$ galaxies are similar in the F200W and F356W bands. 
While for galaxies at $z=3$ and $z=4$ we are probing rest-frame optical morphology, 
for galaxies at $z=5$ and $z=6$ this could be explained by the fact that they present very young stellar populations.
\item At all masses, higher-$z$ galaxies are smaller (more compact) than lower-$z$ ones.
\item There is a mismatch in the mass and light distribution for the more massive galaxies,
more evident at lower redshifts, since $r_{\rm e} > R_{\rm half, \star}$.
Massive galaxies are more compact in mass than in observable stellar light, suggesting that their
light distribution is not well-modeled with a single component
and that they could be (heavily) obscured by dust.
This points to the fact that there could be a morphological transition at $z=4-5$,
which is responsible to build up the first generation of spheroids at $z\gg3$, which
could evolve in bulge+disk systems at lower redshift.
\end{itemize}

This work, and the companion work described in Vega-Ferrero et al.~(\emph{in prep.})
based on forward modeling of simulation data and their analysis using both classical and neural network techniques, 
will be extremely valuable in making a fair comparison between observations and predictions from cosmological simulations, 
as well as interpreting the formation and evolution of galaxies in the mostly unexplored high-$z$ regime.
All data produced in this paper and related to the synthetic images of TNG50 galaxies 
for JWST-like NIRCam and MIRI observations is publicly available at \url{https://www.tng-project.org/costantin22}.


\begin{acknowledgments}

We would like to thank the anonymous referee for improving the content of the manuscript.
L.C. wishes to thank Cristina Cabello, Michele Perna, and William Roper for the useful discussion.

This research has been funded by grant No.~PGC2018-093499-B-I00 and MDM-2017-0737 
Unidad de Excelencia ‘‘Maria de Maeztu’’-Centro de Astrobiolog\'ia (INTA-CSIC) by the Spanish 
Ministry of Science and Innovation/State Agency of Research MCIN/AEI/ 10.13039/501100011033 
and by “ERDF A way of making Europe”.
LC acknowledges financial support from Comunidad de Madrid under 
Atracci\'on de Talento grant 2018-T2/TIC-11612.
AY is supported by an appointment to the NASA Postdoctoral Program (NPP) 
at NASA Goddard Space Flight Center, administered by Oak Ridge Associated Universities 
under contract with NASA.

\end{acknowledgments}


\clearpage

\appendix

\section{Angular resolution effects \label{sec:appendixA}}

In this Appendix we quantified if/how different diagnostics are sensitive to the different 
angular resolutions of the observations. Unfortunately, to date this task
was not carried out for high-$z$ galaxies, since the restframe optical morphology was
inaccessible (at sufficient spatial resolution) with current facilities. 

We focused on $z=3$ galaxies and compared the morphological measurements in the F200W band with angular scale 
of 0.031 arcsec px$^{-1}$ (nominal) with those at 0.015 arcsec px$^{-1}$ (half-nominal) and also 
those in the F356W band at 0.063 arcsec px$^{-1}$ (nominal) with those at 0.030 arcsec px$^{-1}$ (half-nominal).
At $z=3$ we are targeting the restframe optical morphology of our galaxies in 
both NIRCam bands (0.50~$\mu$m in the F200W and 0.89~$\mu$m in the F356W),
thus assuring only mild variation of morphology across wavelength. 
It is worth noting that we are actually probing the effect of drizzling NIRCam images for increasing (doubling) their spatial sampling,
the ordinary strategy that it is foreseen to be adopted by forthcoming JWST observations.
This task is of extreme importance in order to characterize how the galaxy morphology changes across wavelength,
where images could be at different angular resolutions (see Sect.~\ref{sec:section44}).

\begin{figure*}[t!]
\centering
\includegraphics[width=0.65\textwidth, trim=0cm 0cm 0cm 0cm , clip=true]{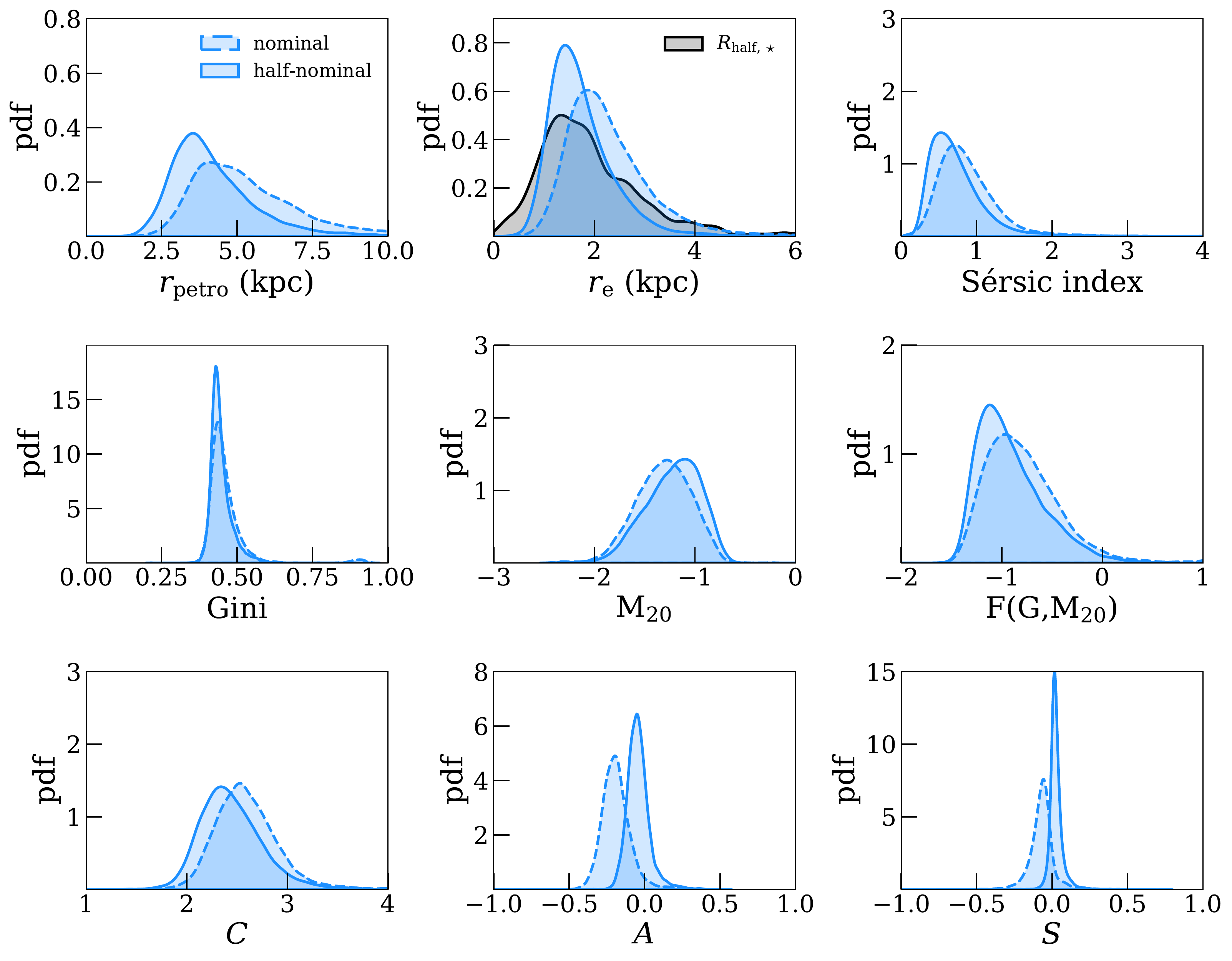}
\caption{Distribution of morphological parameters measured in the F200W band. 
Top panel (from left to right): $r_{\rm petro}$,  $r_{\rm e}$, and S\'ersic index.
Middle panel (from left to right): Gini, M$_{20}$, and F(G,M$_{20}$).
Bottom panel (from left to right): concentration, asymmetry, and smoothness. 
The dashed lines mark the distributions at nominal angular resolution, while the solid lines define the distributions
at half-nominal angular resolution. The distribution of $r_{\rm e}$ (top middle panel) can be compared with the 
one of $R_{\rm half, \star}$ (black curve).
\label{fig:figure9}}
\end{figure*}

\begin{figure*}[t!]
\centering
\includegraphics[width=0.65\textwidth, trim=0cm 0cm 0cm 0cm , clip=true]{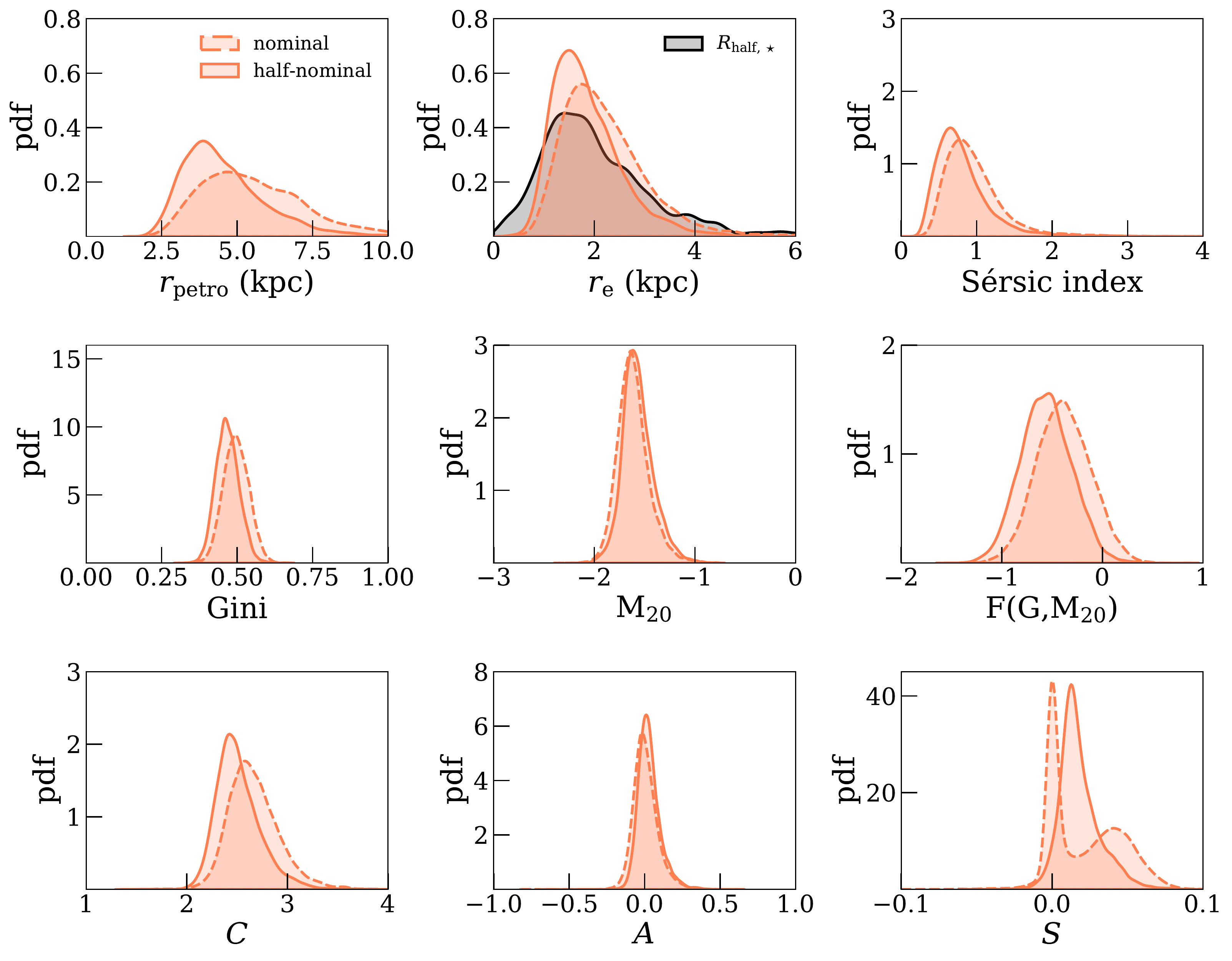}
\caption{As in Fig.~\ref{fig:figure9}, but for the F356W band.
\label{fig:figure10}}
\end{figure*}

In the following, we focused on the following parameters: sizes ($r_{\rm petro}$ and $r_{\rm e}$),
S\'ersic index, Gini, M$_{20}$, F(G, M$_{20}$), and $CAS$ statistics. 
For the rest of the analysis, when referring to sizes, we will use the values measured on elliptical apertures.
In Figs.~\ref{fig:figure9} and \ref{fig:figure10}, we first compare the difference between each parameter 
measured from images at nominal and at half-nominal angular resolution in both filters,
while in Table~\ref{tab:table6}, we reported the median values of each parameter.

In general, we see that sizes do depend on the angular resolution of the images.
Both the Petrosian and the effective radius vary $\sim25\%$ when measured in the F200W and $\sim15\%$ when measured in the F356W band.
We found that the median size is systematically larger if measured on images at nominal resolution, 
even if the values are within the $1\sigma$ variation. 
In the top middle panel of Fig.~\ref{fig:figure9} we see that, at 2~$\mu$m,
the distribution of the half-mass radius almost overlaps with the one of the effective radius measured on
images at half-nominal angular resolution. In the F200W band, the median value of $r_{\rm e}$/$R_{\rm half, \star} = 1.22^{+0.65}_{-0.35}$ 
at nominal and $r_{\rm e}$/$R_{\rm half, \star} = 1.00^{+0.53}_{-0.36}$ at half-nominal angular resolution.
At 3.56~$\mu$m (Fig.~\ref{fig:figure10}), $r_{\rm e}$ measured on
images at half-nominal angular resolution still describes the half-mass radius better than the one at nominal angular resolution,
even though both light-weighted measurements are not probing the tail of small sizes of our galaxies.
In the F356W band, the median value of $r_{\rm e}$/$R_{\rm half, \star} = 1.10^{+0.62}_{-0.30}$ 
at nominal and $r_{\rm e}$/$R_{\rm half, \star} = 0.96^{+0.51}_{0.32}$ at half-nominal angular resolution.
 
The S\'ersic index of the galaxies, being coupled with the effective radius, show a similar trend.
The Gini coefficient is almost insensitive to angular resolution effects ($1-5\%$ variation),
while M$_{20}$ and F(G,M$_{20}$) do slightly depend on the angular sampling of the images 
($\sim5-10\%$ variation for M$_{20}$ and $\sim10-25\%$ variation for F(G,M$_{20}$)).
The values are compatible within $1\sigma$ variation.
Following a similar approach to the one applied in this work, 
\citet{Bignone2020} explored the impact of measuring the optical morphology of
$z=0.1$ galaxies in the EAGLE simulations \citep{Schaye2015, Crain2015, McAlpine2016} at different angular resolution.
They found that the Gini coefficient systematically reduces with decreasing spatial resolution,
while M$_{20}$ was almost not affected (median shifts less than 0.8 percent).

Finally, the $CAS$ statistics seem to be not well-suited to characterize the morphology
of high-$z$ galaxies in our sample. Among the three diagnostics, 
the galaxy concentration is the only one that provides sensible results, 
being almost independent of the drizzling effect.

\begin{deluxetable}{cDDcDDc}
\tablecaption{Morphological parameters measured in the F200W and F356W bands at different angular resolutions for galaxies at $z=3$. 
\label{tab:table6}}
\tablehead{
\colhead{Diagnostic} &\multicolumn2c{Median value} &\multicolumn2c{Median value} & Variation &\multicolumn2c{Median value} &\multicolumn2c{Median value} & Variation \\
\colhead{} & \multicolumn2c{(F200W nominal)} & \multicolumn2c{(F200W half-nominal)} & (\%) & \multicolumn2c{(F356W nominal)} & \multicolumn2c{(F356W half-nominal)} & (\%)
}
\decimalcolnumbers
\phd
\startdata
$r_{\rm petro}$ (kpc) 	 & 	$4.97^{+2.03}_{-1.26}$	&	$3.90^{+1.51}_{-0.92}$	   &	24	& 	$5.37^{+1.97}_{-1.48}$	&	$4.30^{+1.59}_{-1.01}$	   &	17	\\
$r_{\rm e}$ (kpc) 		& 	$2.11^{+0.88}_{-0.57}$	&	$1.61^{+0.69}_{-0.43}$	   &	25	& 	$2.06^{+0.89}_{-0.62}$	&	$1.72^{+0.76}_{-0.51}$	   &	16	\\
S\'ersic index	 		&	$0.80^{+0.39}_{-0.25}$	&	$0.65^{+0.37}_{-0.24}$   	   &	24	&	$0.89^{+0.37}_{-0.26}$	&	$0.73^{+0.36}_{-0.23}$	   &	20	\\
Gini		 			&	$0.45^{+0.04}_{-0.02}$	&	$0.44^{+0.04}_{-0.02}$   	   &	1	&	$0.50^{+0.04}_{-0.04}$	&	$0.47^{+0.04}_{-0.04}$	   &	5	\\
M$_{20}$				&	$-1.30^{+0.27}_{-0.28}$	&	$-1.19^{+0.24}_{-0.32}$   	   &	9	&	$-1.63^{+0.16}_{-0.13}$	&	$-1.58^{+0.16}_{-0.12}$	   &	3	\\
F(G, M$_{20}$) 		&	$-0.84^{+0.40}_{-0.29}$	&	$-0.98^{+0.40}_{-0.24}$   	   &	-11	&	$-0.39^{+0.28}_{-0.26}$	&	$-0.56^{+0.26}_{-0.25}$	   &	-26	\\
Concentration	 		&	$2.56^{+0.31}_{-0.26}$	&	$2.41^{+0.32}_{-0.26}$   	   &	6	& 	$2.64^{+0.26}_{-0.21}$	&	$2.48^{+0.24}_{-0.17}$	   &	6	\\
Asymmetry	 		& 	$-0.20^{+0.09}_{-0.08}$	&	$-0.05^{+0.07}_{-0.06}$	   &	151	& 	$-0.01^{+0.08}_{-0.06}$	&	$0.02^{+0.07}_{-0.06}$	   &	-100	\\
Smoothness	 		& 	$-0.07^{+0.04}_{-0.06}$	&	$0.02^{+0.03}_{-0.02}$	   &	-262	& 	$0.02^{+0.03}_{-0.02}$	&	$0.01^{+0.02}_{-0.01}$	   &	100	\\
\enddata
\tablecomments{
Values could slightly differ from those in other tables, because of the quality criteria explained in Sect.~\ref{sec:section41}.
(1) Morphological parameter. 
(2) Median value and 16th-84th percentile range at nominal angular resolution (0.031 arcsec px$^{-1}$)  in the F200W band.
(3) Median value and 16th-84th percentile range at half-nominal angular resolution (0.015 arcsec px$^{-1}$) in the F200W band.
(4) Percentage of the variation between parameters at different angular resolutions.
(5) Median value and 16th-84th percentile range at nominal angular resolution (0.063 arcsec px$^{-1}$) in the F356W band.
(6) Median value and 16th-84th percentile range at half-nominal angular resolution (0.030 arcsec px$^{-1}$) in the F356W band.
(7) Percentage of the variation between parameters at different angular resolutions.
}
\end{deluxetable}

\clearpage

\section{PSF effects: $r_{\rm half, phot}$ and $r_{\rm e}$ \label{sec:appendixB}}

\begin{figure*}[t!]
\centering
\includegraphics[width=0.75\textwidth, trim=0cm 0cm 0cm 0cm , clip=true]{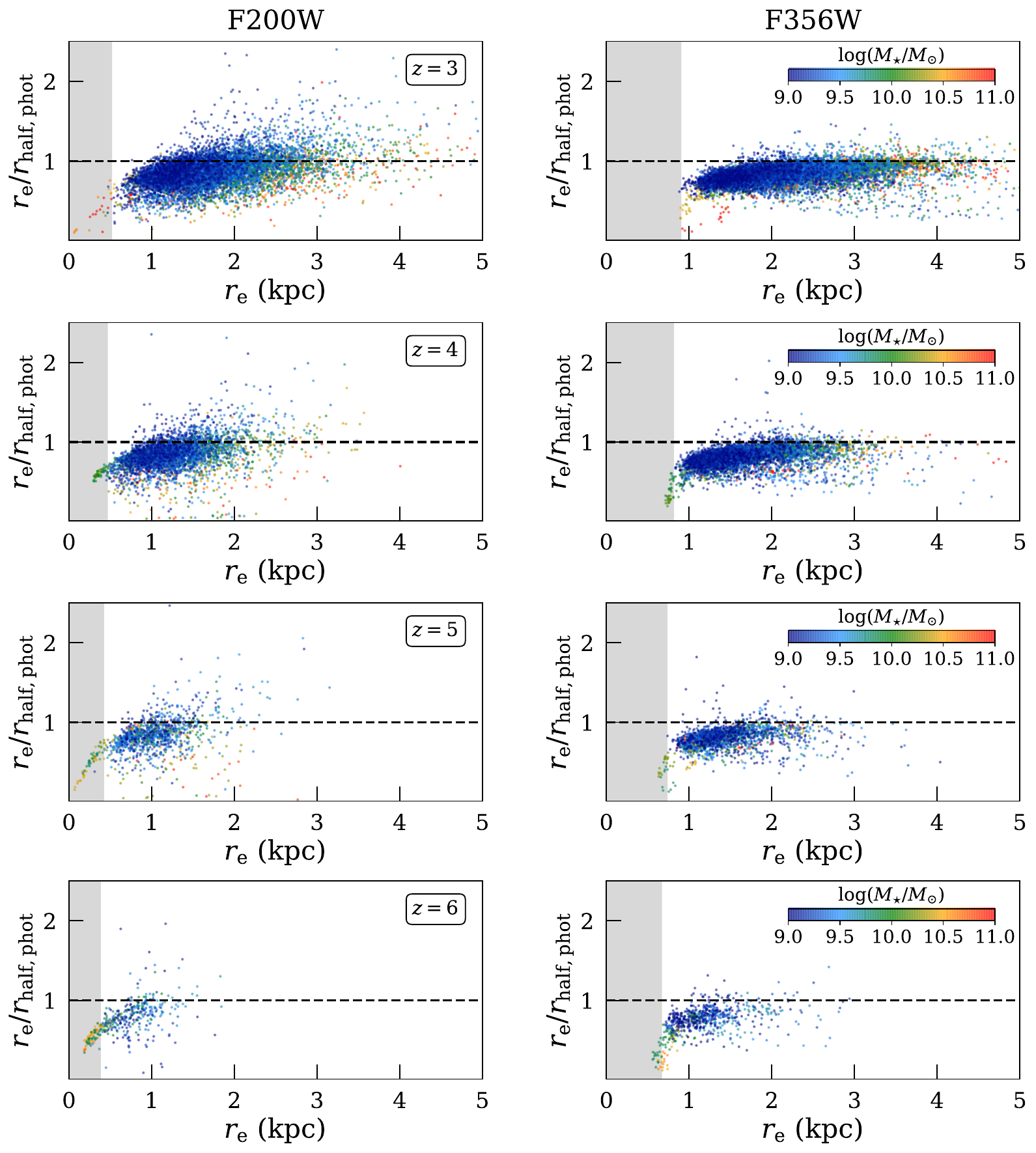}
\caption{Comparison between $r_{\rm half, phot}$ and $r_{\rm e}$ for galaxies at different redshift (from top to bottom)
in the F200W band (left panels) and F356W band (right panels).
Data points are color-coded according to the value of the galaxies' total stellar mass.
The black dashed line marks the 1:1 relation of the ratio $r_{\rm e}$/$r_{\rm half, phot}$.
The gray shaded regions mark the size of the full width at half maximum of the PSF at each redshift.
\label{fig:figure11}}
\end{figure*}

In this Appendix, we compared the role of the half-light radius and the effective radius in
probing the size of our galaxies. Indeed, there is a meaningful difference in the two measurements,
since $r_{\rm half, phot}$ does not take into account the effect of the PSF, while $r_{\rm e}$ does.
Before running \texttt{statmorph} on the drizzled images (Sect.~\ref{sec:section41}) and
describing the galaxy light with a S\'ersic model, we used the three stars in the NIRCam field-of-view 
and the \texttt{photutils.psf} package \citep{Bradley2020} to built an observed PSF model \citep[e.g.,][]{Cabello2022} at half-nominal angular resolution.
Thus, we measured the galaxy size and compared the ratio between $r_{\rm half, phot}$ with $r_{\rm e}$ at different redshifts. 
In Fig.~\ref{fig:figure11} we show how the trend between the two radii changes and
how the $r_{\rm half, phot}$ and $r_{\rm e}$ distributions compare with the distribution of the half-mass radius.
In particular, while the median value of $r_{\rm half, phot}$ is compatible with the median value of $R_{\rm half, \star}$, 
we can appreciate in Fig.~\ref{fig:figure11} how $r_{\rm half, phot}$ is not able to describe the size of the smallest galaxies
and the ratio $r_{\rm half, phot} / r_{\rm e}$ increasingly deviates from the 1:1 relation at smaller radii.

\clearpage

\section{1D surface-brightness profiles: examples \label{sec:appendixC}}

In this Appendix, we provided some examples of the one-dimensional surface-brightness profiles of our galaxies ($z=3$, F356W).
We measured the ellipse-averaged radial profiles of surface brightness both in the observed image and in 
the seeing-convolved model \citep{Costantin2018b, Costantin2018a}.
We used the \texttt{isophote.Ellipse.fit\_image} in Photutils from Python’s astropy package \citep{Bradley2020}.

\begin{figure*}[t!]
\centering
\includegraphics[width=0.75\textwidth, trim=0cm 0cm 0cm 0cm , clip=true]{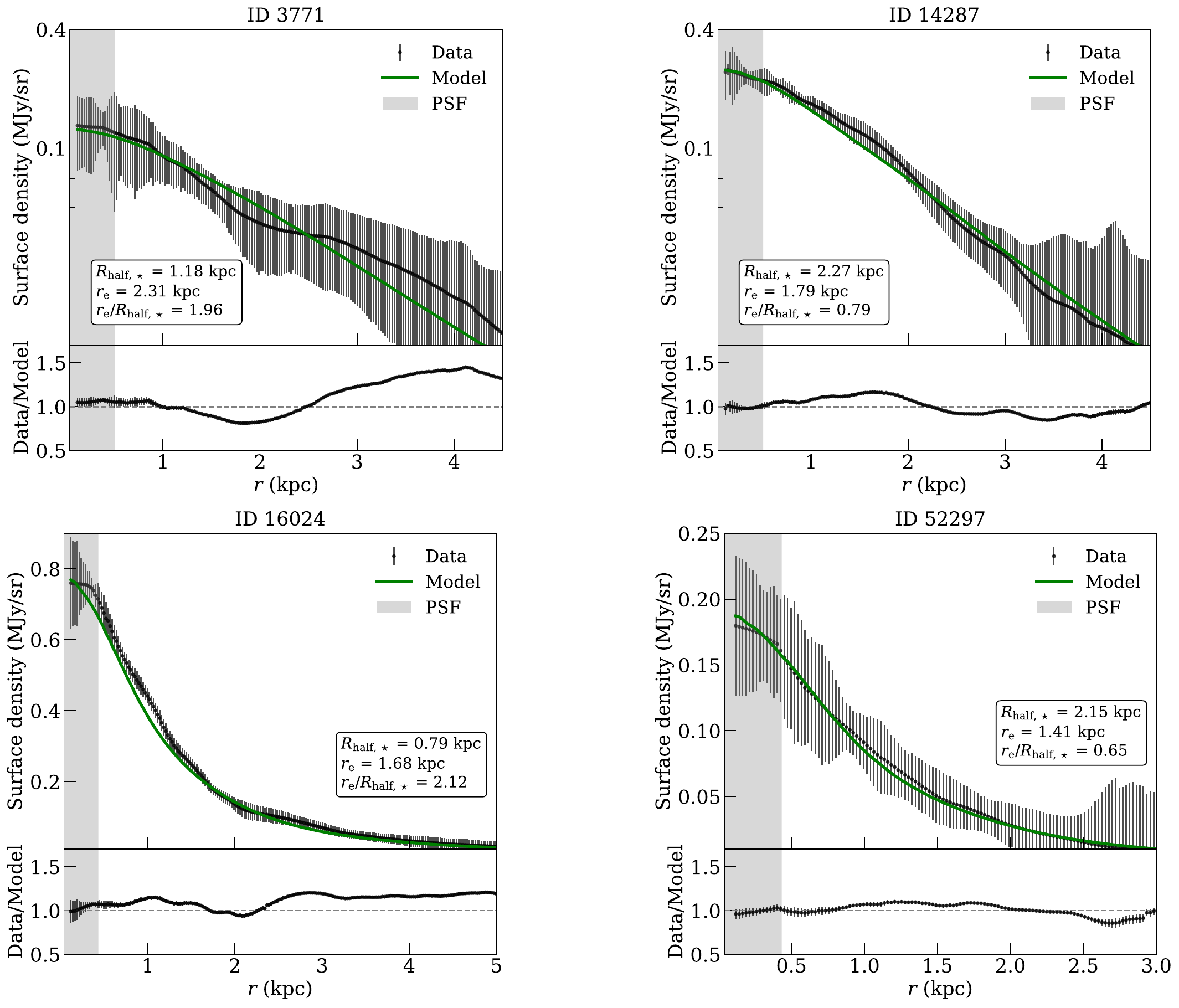}
\caption{Ellipse-averaged radial profiles of surface brightness measured in the observed (black dots with grey error bars) 
and seeing-convolved modeled images (green solid line) and their corresponding difference.
\label{fig:figure12}}
\end{figure*}

In Fig.~\ref{fig:figure12}, we show two galaxies with $r_{\rm e}$/$R_{\rm half, \star} < 1$ (ID 14287, ID 52297) 
and two galaxies with $r_{\rm e}$/$R_{\rm half, \star} > 1$  (ID 3771, ID 16024).
From this comparison, we see that the one-dimensional profiles of our S\'ersic model seem reasonable, 
but the ratios $r_{\rm e}$/$R_{\rm half, \star}$ deviate from unity anyway. Thus, to fully constrain the complexity 
of these galaxies, we release both their images and the morphological catalog with every physical parameter.
In this way, it is possible to properly derive the two-dimensional model and carefully study each individual object.

\clearpage


\bibliographystyle{aasjournal}

\allauthors

\end{document}